\documentclass[conference, 10pt]{IEEEtran}

\ifCLASSINFOpdf
\else
\fi

\usepackage{amsmath}

\usepackage{authblk}

\usepackage[colorinlistoftodos,prependcaption,disable]{todonotes} %

\usepackage{hyperref}
\usepackage{xurl}
\usepackage{cite}
\usepackage{booktabs}
\PassOptionsToPackage{hyphens}{url}
\usepackage{url}
\usepackage{multirow}
\usepackage{svg}
\usepackage{adjustbox}
\usepackage[table]{xcolor}
\newcommand{\rot}[1]{\multicolumn{1}{c}{\adjustbox{angle=60,lap=\width-1em}{#1}}}

\newcommand{\parvspace}{\vspace{0.02cm}}

\RequirePackage{fontawesome}

\begin{document}

\title{Hey there! You are using WhatsApp:\\
Enumerating Three Billion Accounts for \\Security and Privacy 
}

\makeatletter
\renewcommand\AB@affilsepx{, \protect\Affilfont}
\setlength{\affilsep}{0.5em}   %
\makeatother

\author[1,2]{Gabriel K. Gegenhuber}
\author[3]{Philipp É. Frenzel}
\author[1]{Maximilian Günther}
\author[1]{Johanna Ullrich}
\author[1]{Aljosha Judmayer}

\affil[1]{University of Vienna, Faculty of Computer Science}
\affil[2]{UniVie Doctoral School Computer Science}
\affil[3]{SBA~Research}

\IEEEoverridecommandlockouts
\makeatletter\def\@IEEEpubidpullup{6.5\baselineskip}\makeatother
\IEEEpubid{\parbox{\columnwidth}{
		Network and Distributed System Security (NDSS) Symposium 2026\\
		23-27 February 2026, San Diego, CA, USA\\
		ISBN 979-8-9919276-8-0\\
		https://dx.doi.org/10.14722/ndss.2026.230805\\
		www.ndss-symposium.org
}
\hspace{\columnsep}\makebox[\columnwidth]{}}

\maketitle

\begin{abstract}
WhatsApp, with 3.5 billion active accounts as of early 2025, is the world's largest instant messaging platform. 
Given its massive user base, WhatsApp plays a critical role in global communication.

To initiate conversations, users must first discover whether their contacts are registered on the platform.
This is achieved by querying WhatsApp's servers with mobile phone numbers extracted from the user’s address book (if they allowed access).
This architecture inherently enables phone number enumeration, as the service must allow legitimate users to query contact availability.
While rate limiting is a standard defense against abuse, we revisit the problem and show that WhatsApp remains highly vulnerable to enumeration at scale.
In our study, we were able to probe over a hundred million phone numbers per hour without encountering blocking or effective rate limiting.

Our findings demonstrate not only the persistence but the severity of this vulnerability.
We further show that nearly half of the phone numbers disclosed in the 2021 Facebook data leak are still active on WhatsApp, underlining the enduring risks associated with such exposures.
Moreover, we were able to perform a census of WhatsApp users, providing a glimpse on the macroscopic insights a large messaging service is able to generate even though the messages themselves are end-to-end encrypted. 
Using the gathered data, we also discovered the re-use of certain X25519 keys across different devices and phone numbers, indicating either insecure (custom) implementations, or fraudulent activity.

In this updated version of the paper, we also provide insights into the collaborative remediation process through which we confirmed that the underlying rate-limiting issue had been resolved.

\end{abstract}

\IEEEpeerreviewmaketitle

\pagestyle{plain}

\section{Introduction}

WhatsApp is the world's largest instant messaging service with 3.5\,B (billion) active accounts as of writing and plays a critical role in global communication. %
Holding this unique position entails a significant responsibility to safeguard its users.
WhatsApp has improved security and privacy measures over the years, most notably the deployment of end-to-end encryption (E2EE)~\cite{noauthor_whatsapp_about}.
With E2EE, messaging service operators can no longer access the content of user messages.
Still, the possibility to construct social graphs of their users by observing message patterns is ingrained in prevalent E2EE 
designs~\cite{kales_mobile_2019}. 
In addition to the threat posed by curious or compromised service operators, malicious users also represent a significant risk, as they may attempt to extract information about other account holders.

WhatsApp maintains a registry of its users linked to their phone numbers to support seamless contact discovery.
When users install the app, they can choose to give permission to access their local address book and upload it to WhatApp's servers.
In return, the user is informed which contacts are registered on the platform. While this mechanism was designed for convenience and represented a key driver of WhatsApp in achieving widespread adoption, it can be misused to check whether a specific individual --- such as a government official, former partner, or employer --- is registered on WhatsApp, thereby making presence information available if you know their phone number.
This basic functionality has previously been exploited at scale.
In 2012, Schrittwieser et al.~\cite{schrittwieser_2012_guess} enumerated 10\,M (million) phone numbers and found 21,095 active WhatsApp accounts in less than 2.5 hours.
Nearly a decade later, Hagen et al.\cite{hagen_2021_all} expanded on this by probing 50.5\,M numbers --- 10\% of all US phone numbers --- discovering 5\,M active accounts over 34 days.
Privacy-preserving approaches for contact discovery have been developed, such as private set intersection (PSI)~\cite{kales_mobile_2019} or based on trusted hardware~\cite{signal2017private}, but their deployment is non-trivial given necessary set sizes and practical constraints~\cite{hagen_2021_all}.

Knowing whether a specific (mobile) phone number is linked to a messaging app is highly sensitive, especially when that number is tied to a known individual. 
In regions where certain messaging apps are banned (e.g., China or Myanmar), such information could carry serious consequences for the user. 
Moreover, large-scale databases of registered phone numbers can be misused by attackers.
Since a registered number typically indicates an active device, these lists are a reliable basis for spam, phishing, or robocall attacks.
This naturally raises the question of how long information, once collected, stays valid and thus can be (ab)used by criminals. 

\begin{figure}[t]
  \centering
  \includegraphics[width=0.8\linewidth,trim={6.3cm 9.7cm 6.3cm 9.5cm},clip]{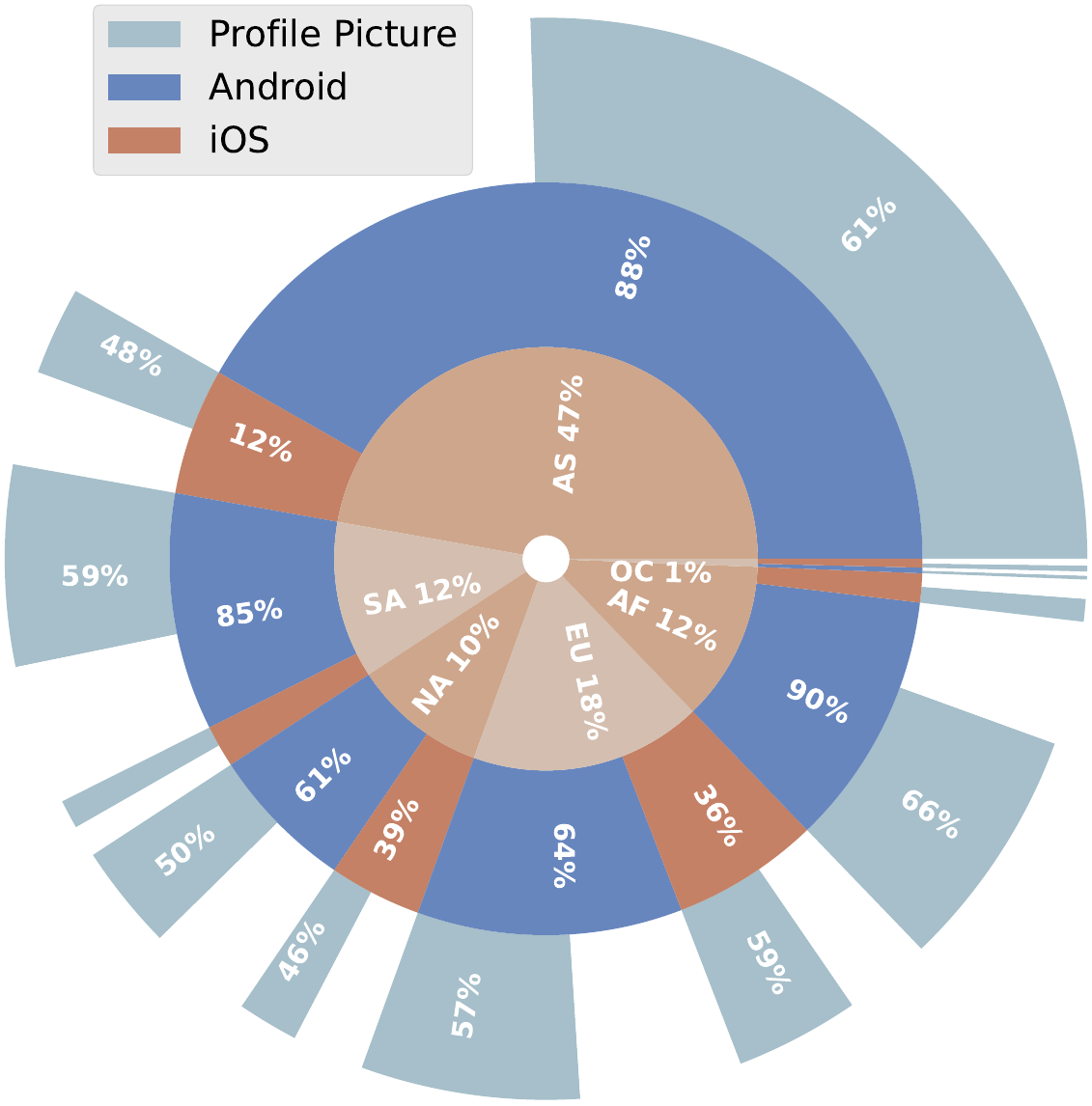}
  \caption{WhatsApp users wrt. to continent, Android vs. iOS use, and profile picture for 3.5\,B users.
  }
  \label{fig:sunburst}
\end{figure}

This paper reassesses the topic of phone number enumeration in WhatsApp, analyzes the extent of information revealed by WhatsApp to malicious users, as well as investigates the permanence of once-released data, and concludes that this threat is not only still present but significantly worse than expected.
We developed a novel method to quickly generate candidate datasets of potentially active phone numbers for 245 countries.
To evaluate our approach, we leveraged available open-source WhatsApp clients to query candidate phone numbers and verify whether they correspond to active WhatsApp accounts.
The combination of automatic phone number generation and direct access via a reverse-engineered WhatsApp API --- instead of the app's user interface or an automation thereof as done by previous work~\cite{hagen_2021_all} ---  enabled exhaustive enumeration of its user base, thereby revealing phone numbers associated with a substantial share of the global population.
Additionally, direct access to WhatsApp's XMPP API allows extraction of more details on its user population.
Following the rationale of Hagen et al.\cite{hagen_2021_all}, we argue that the only reliable way to assess the feasibility and effectiveness of phone number enumeration attacks on instant messaging platforms is through empirical evaluation --- \emph{``the only way to reliably estimate the success of our attacks in the real world''}~\cite{hagen_2021_all}.

As we assumed that our query attempts, always originating from the same IP address and the same five accounts, will get rate-limited and blocked swiftly, we initially focused our evaluation on the US to measure how many of the phone numbers leaked in 2019~\cite{facebook_leak_2021} are still active. 
To our surprise, neither our IP address nor our accounts have been blocked by WhatsApp\footnote{Apart from one time-limited ban induced by an error on our side.}. 
Moreover, we did not experience any prohibitive rate-limiting. 
With our query rate of 7,000 phone numbers per second (and session), we could confirm 3.5\,B phone numbers registered on WhatsApp (exceeding the ``more than 2 billion people'' officially stated by WhatsApp\cite{noauthor_whatsapp_about}).
Beyond the existence of the phone number and its associated device list, we were also able to obtain data for each discovered account
consisting of: Public keys for E2EE encryption, meta information such as timestamps\footnote{Timestamps for key updates, profile picture, and about text.} and, if available and shared by the user, profile picture, about text and business account information\footnote{See Section~\ref{sec:dataset} for detailed list of the retrieved data.}.

While most data points may seem harmless, especially in isolation, their large-scale aggregation can provide meaningful and potentially revealing insights.
The sheer quantity of data points allows us not only to perform a detailed census of the WhatsApp user population but also to provide a glimpse of the interesting macroscopic insights a large-scale instant messaging service is able to obtain, even without having access to message content.
For example, Figure~\ref{fig:sunburst} shows the share of Android vs. iOS devices per continent.
The figure shows the higher prevalence of iOS in economically stronger regions and diverging shares of available profile pictures, indicating regional and cultural differences regarding privacy norms.

Comparing our retrieved dataset of 3.5\,B records (i.e., active accounts) to existing scraped datasets, it surpasses the dataset of the 2021 Facebook scraping incident including around 500\,M records by a factor of five\footnote{
As a point of reference in terms of size, the currently largest data breach happened 2013 at Yahoo~\cite{Wikipedia_Yahoo_data_breaches} and affected around 3\,B accounts.}.
The fact that we could obtain this data unhindered allows for the possibility that others may have already done so as well.
Beyond this, our contributions are as follows:

\begin{itemize}
    \item We developed a method to generate plausible mobile phone numbers for 245 countries. Using this method, we can shrink the candidate set of potentially assigned mobile phone numbers worldwide to 63\,B.
    \item With 3.5\,B records (i.e., active accounts), we analyze a dataset that would, to our knowledge, classify as the largest data leak in history, had it not been collated as part of a responsibly-conducted research study. 
    The dataset contains phone numbers, timestamps, about text, profile pictures, and public keys for E2EE encryption, and its release would entail adverse implications to the included users.
    \item We compare our dataset with the publicly available data of the 2021 Facebook data scraping incident, with 500\,M entries. Half of them are still relevant and active, even six years after the original scraping. This highlights the long-lasting consequences of data incidents.
    \item For ethical reasons, we refrain from analyzing individual users. Instead, when analyzing this dataset, we perform a census on the WhatsApp population to show the wealth of possibilities.
    Among others, we provide macroscopic insights on WhatsApp accounts per country and capita, growth and churn, the availability of about text, profile picture, and companion devices, operating system (OS) shares (e.g., Android vs. iOS), account activeness, device age, and %
    popularity.
    The census also emphasizes the far-reaching insight a messaging platform like WhatsApp still has despite E2EE.
    \item We highlight the potential consequences of phone number enumeration for users in countries where WhatsApp is banned, such as China, Myanmar and North Korea, by identifying active numbers in those countries. 
    Coincidentally, we were also able to show that even before lifting the ban on WhatsApp in Iran, a significant portion of active accounts had already existed, indicating that the ban was ineffective. 
    \item To the best of our knowledge, the dataset is also one of the largest of cryptographic (X25519) public keys generated and deployed in the field. 
    By analyzing these keys, we were able to detect the re-occurrence of public keys across different accounts, 
    We discovered extensive re-use (multiple hundred times) of certain public keys, as well as the repeated occurrence of one-time prekeys across different devices, both not easily explained with benign uses-cases, indicating problems in custom implementations or fraudulent activity.  
    As an example, 20 phone numbers (mainly in the US) use an all-zero private key, suggesting a broken pseudo-random number generator in this case or the use of non-standard software. 

\end{itemize}

The remainder of the paper is organized as follows: 
Section~\ref{sec:methodology} presents our phone number enumeration methodology, which was developed from an initial exploration with limited scope into a large-scale measurement with global coverage, and its application to the WhatsApp messaging platform, eventually collecting a dataset of 3.5\,B accounts.
Section~\ref{resultsusers} evaluates the collected dataset to gain insight on the WhatsApp population, among others, the deployment in different countries, including those in which WhatsApp is banned, the availability of about text and profile pictures, and account churn.
Section~\ref{sec:resultsprekeys} and Section~\ref{sec:prekey_bundle_analysis} evaluate the E2EE keying material (i.e., public keys and key IDs) of the 3.5\,B accounts, revealing further insights, including OS shares, device age, account activeness and key reusage.
Comparing the collected dataset from WhatsApp with the 2021 Facebook scraping incident, Section~\ref{sec:comparison-facebook-leak} clarifies how long information once leaked remains valid and could be abused by adversaries.
Section~\ref{sec:relatedwork} elaborates related work, Section~\ref{sec:discussion} discusses our results, and Section~\ref{sec:conclusion} concludes.
Finally, we consider ethics and responsible disclosure.

\section{Testing Methodology and Implementation}
\label{sec:methodology}

Reassessing the feasibility of phone number enumeration, we started with a careful initial exploration.
Contrary to our expectations, we did not encounter prohibitive rate-limiting or any contact attempts by the platform operator.
Consequently, we further developed our method for global coverage.
Using WhatsApp's XMPP APIs, we extracted about text and profile picture, when allowed by privacy controls, as well as public keys and timestamps used for E2EE for all discovered WhatsApp users.
The measurements were conducted in multiple rounds between the middle of December 2024 and the middle of April 2025.
We wish to highlight that we made no effort to disguise our activity --- all queries were sent from a single server at our university.
This section provides further details on our methodology and implementation.

\subsection{Initial Exploration}
Our initial exploration aimed to understand WhatsApp endpoints for potential data collection and executed small-scale experiments on user enumeration with astonishing results.

\begin{table}
\centering
\newcommand\tH{\cellcolor{lightgray!30}}
\newcommand\tFa{$^{\ast}$}
\newcommand\tFb{$^{\dag}$}
\begin{adjustbox}{width=1\columnwidth}
\begin{tabular}{>{\kern-\tabcolsep}lllr<{\kern-\tabcolsep}} %
\toprule
Protocol & API Endpoint    & Returned Infos                           & Speed\tFb \\ \midrule
\multirow{7}{*}{XMPP} 
& IsOnWhatsapp                        & true / false                           & 7,000                          \\ %
& \tH                                 & Device Indexes (Main \& Companion \tH  & \tH                            \\ %
& \multirow{-2}{*}{GetDeviceList} \tH & Devices) \tH                           & \multirow{-2}{*}{7,000} \tH    \\
& \multirow{2}{*}{GetUserInfo}        & Profile Picture (URL, TS), About Text  & \multirow{2}{*}{3,000}         \\ %
&                                     & (Content, TS), Business Info (Name)    &                                \\
& \tH                                 & Identity Key, Signed Prekey, \tH       & \tH                            \\ %
& \multirow{-2}{*}{GetPrekeys} \tH    & One-Time Prekey, TS \tH                & \multirow{-2}{*}{2,000} \tH    \\ \arrayrulecolor{black!40}\midrule
HTTP & FetchPicture                   & Profile Picture (High Quality)         & 5,500                          \\ \bottomrule
\end{tabular}
\end{adjustbox}
\\
\vspace{1ex}
TS~Timestamp \hspace{1ex}
\tFb~Approx. queried accounts per second. \\
\vspace{1ex}
\caption{Overview of WhatsApp Endpoints and Enumeration Speed (as Experienced in Measurements).}
\label{tab:endpoint-overview}
\end{table}

\parvspace
\noindent
\textbf{API-Level WhatsApp Access.}
Previous work~\cite{schrittwieser_2012_guess, hagen_2021_all} used the official WhatsApp client for phone number enumeration and were consequently limited to its regular user interface.
In contrast, this work leverages unofficial implementations to directly query WhatsApp's XMPP endpoints, see Table~\ref{tab:endpoint-overview}, returning more detailed information.
For most of the crawling, we based our work on \textit{whatsmeow}\footnote{\url{https://github.com/tulir/whatsmeow}}, a community-driven open-source client that is based on the findings of reverse engineering  WhatsApp's official web client.

Our custom implementation allows querying whether a phone number is registered on WhatsApp.
It is also possible to retrieve registered numbers' about text and profile picture, if available, and a list of companion devices (i.e., secondary devices like tablets or WhatsApp Web registered with the same account).
For all registered devices, it is possible to query the public keys used for E2EE.
Batching, i.e., cumulating multiple accounts, is possible for most queries, significantly reducing their overall number.
The client requires an authenticated session, which can be established by scanning a QR code to pair with an active WhatsApp account.
A maximum of five concurrent sessions are allowed per WhatsApp account.

\parvspace
\noindent
\textbf{Exploratory Input and Setup.}
Anticipating rate limits and other protection mechanisms from WhatsApp, we began with smaller measurements, extending them in successive steps.
Initially, we enumerated all valid US phone numbers from the 2021 Facebook data leak and finished in less than an hour.
Then, we enumerated all valid phone numbers in this dataset in less than a day.
For phone number validation and country assignment, we relied on Google's widely used open-source library \textit{libephonenumber}.
For data retrieval, we used a single server at our university, limited ourselves to five concurrent sessions, and batched 50K accounts in one query to avoid overstressing WhatsApp's infrastructure.
Throughout the study, we ensured reachability by providing an abuse email address for the server's IP address.

In summary, it was not only possible to enumerate all 488M phone numbers from the Facebook data leak, but it was also possible at high speed.
We could infer their registration and current about text, profile picture, companion devices, and E2EE-related public keys.
Table~\ref{tab:endpoint-overview} provides the number of accounts enumerated per second.
We encountered no rate limits, our accounts were not banned from the platform, and we received no complaints via email. 
For a detailed evaluation of the phone numbers included in the 2021 Facebook data leak, we refer to Section~\ref{sec:comparison-facebook-leak}.

\subsection{Global Phone Number Generation}
Given the possibility of phone number enumeration, the achieved rate, and the lack of any response from WhatsApp, the idea of systematically enumerating phone numbers worldwide arose after a week of experimentation.
However, the E.164 phone number format~\cite{itu_e164} allows up to 15 digits, rendering exhaustive enumeration --- even at the fastest rates --- infeasible.

\begin{table}[tb]
\centering
\definecolor{highlight}{RGB}{255,229,229}
\newcommand\tH{\cellcolor{highlight}}
\begin{tabular} {>{\kern-\tabcolsep}rlrr<{\kern-\tabcolsep}} %
\toprule
Code   & Country     & \# Candidates  & \# Candidates (pp)   \\ \midrule
+43            & AT          & 511.11\,B \tH  &  0.48\,B \tH  \\
+62            & ID          &  88.88\,B \tH  &  4.44\,B \tH  \\
+55            & BR          &   7.37\,B \tH  & 14.74\,B \tH  \\
+86            & CN          &   5.21\,B      &  5.21\,B      \\
+52            & MX          &   4.48\,B \tH  &  8.96\,B \tH  \\
+91            & IN          &   3.33\,B      &  3.33\,B      \\
+1             & US          &   2.87\,B      &  2.87\,B      \\
+49            & DE          &   1.33\,B      &  1.33\,B      \\
+880           & BD          &   1.17\,B      &  1.17\,B      \\
+358           & FI          &   1.11\,B      &  1.11\,B      \\
+39            & IT          &   1.08\,B      &  1.08\,B      \\
+7             & RU          &   1.00\,B      &  1.00\,B      \\ %
\multicolumn{2}{r}{Residual} &  17.52\,B      & 17.52\,B      \\ \midrule
\multicolumn{2}{r}{\textbf{TOTAL}}    & 646.39\,B \tH  & 63.17\,B \tH  \\ \bottomrule
\end{tabular}
\\
\vspace{1ex}
\caption{\textit{libphonegen} yields 63.17\,B phone numbers for enumeration. 
\colorbox{highlight}{Highlighted countries} required post-processing (pp). }
\label{tab:libphoennumber-candidates}
\end{table}

\parvspace
\noindent
\textbf{\textit{libphonegen.}} For search-space reduction, we have thus explored and compared different approaches:
\begin{itemize}
    \item The International Telecommunication Union (ITU) defined numbering plans for its member countries~\cite{itu_countrycodes}. While significantly reducing the search space (e.g., by reducing the number of digits for individual countries), the plans primarily document allowed allocations rather than actual assignments. 
    In practice, utilizing these plans to generate phone numbers could trigger alerts at the targeted service, as many of these numbers are unused.
    Furthermore, the plans are published in non-uniform document formats (MS Word, PDF), requiring manual extraction of relevant information for each country.
    \item By scraping public services, such as online phone books, or leaked data, similar to the \textit{rockyou.txt} dictionary for passwords~\cite{rockyou_wikipedia}, a phone number 'hitlist' could be generated. On the positive side (for an attacker), these sources often include metadata, e.g., names, email addresses, or employers, that could be linked with the discovered WhatsApp accounts. On the negative side, a hitlist only covers a share of the active accounts.
    \item Google's \textit{libphonenumber}\footnote{\url{https://github.com/google/libphonenumber}} is an actively maintained library that is widely used for parsing, formatting, and validating phone numbers. This ensures an up-to-date view of valid phone numbers. It also provides additional metadata (e.g., operator, or landline vs. mobile number) that is helpful for the validation (or generation) of input sets.
\end{itemize}

 Eventually, we developed \textit{libphonegen}, a phone number generator leveraging the existing \textit{libphonenumber's} national number format data, and generated 646\,B mobile phone numbers for the 245 ISO~3166-1 countries~\cite{ISO3166-1}.
Table~\ref{tab:libphoennumber-candidates} shows all countries with more than 1\,B candidate numbers.

\parvspace
\noindent
\textbf{Post-Processing (AT, IN).}
As apparent in Table~\ref{tab:libphoennumber-candidates}, Austria (AT) and Indonesia (ID) have the highest numbers of phone numbers, accounting together for 93\% of the total number space.
In contrast to most countries, Austria does not specify a fixed amount of numbers for mobile networks.
Anything between 7 and 13 digits is possible, significantly increasing the search space.
However, only specific number blocks and shorter numbers are used in practice.
To refine our candidate set, we applied a hitlist-based approach to identify popular number ranges, reducing the respective amount to 480\,M numbers.
For Indonesia, we reduced the search space to numbers with specific operator prefixes found on Wikipedia and posts on Indonesian web forums,
effectively yielding 4.44\,B phone numbers in total for enumeration.
We enumerated the complete space for the other countries as proposed by \textit{libphonenumber}.

\parvspace
\noindent
\textbf{Post-Processing (BR, MX).}
Our measurements revealed an unexpectedly low number of accounts per capita in Brazil (BR) and Mexico (MX).
We found that both countries had recently modified their mobile phone number formats~\cite{fcc_mexico_dialing} and \textit{libphonenumber} no longer recognizes the old formats.
Yet, it appears that many WhatsApp accounts are still registered with these legacy numbers and remain discoverable only when querying this old format~\cite{zoko_whatsapp_id}.
Consequently, we enhanced our input dataset by these numbers, see again Table~\ref{tab:libphoennumber-candidates}.

\subsection{Measurement with Global Coverage}
After phone number generation with \textit{libphonegen} and post-processing for four countries, we ended up with 63\,B phone numbers for 245 countries and enumerated them on WhatsApp.
The measurement setup was equivalent to that in the exploratory phase, i.e., a single server limited to five concurrent sessions and batching 50K accounts in one query.
In a first round, we generally checked for a number's registration at WhatsApp using \textit{GetDeviceList}, see Table~\ref{tab:endpoint-overview} for an overview of the different endpoints.
In a second round, we retrieved account information such as current public about text and profile picture using \textit{GetUserInfo}, and in a final round, the associated public keys, key IDs and timestamps using \textit{GetPrekeys}.
The measurement rounds were conducted between mid-December 2024 and mid-April 2025.

\begin{figure}[t]
    \centering
    \includegraphics[width=0.8\linewidth]{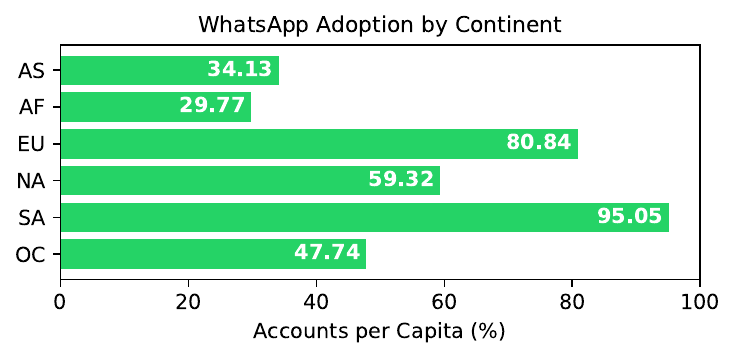}
    \vspace{-0.3cm}
    \caption{WhatsApp adoption per continent. Percentage shares are calculated by dividing the number of discovered WhatsApp accounts by the respective population size (per capita) of each continent.}
    \label{fig:market-penetration-continent-barchart}
    \vspace{-0.2cm}
\end{figure}

\parvspace
\noindent
\textbf{Limitations.} 
Due to data collection in multiple rounds, the resulting dataset has sporadic blind spots, as accounts might have been deleted meanwhile and it was not possible anymore to request additional data for them.
We discovered 3,456,622,387 unique phone numbers (accounts) across 235 countries in the first round, but could not collect metadata, such as profile about text or keying materials, for all of them in the later rounds.
Additionally, we conducted complementary measurements to investigate WhatsApp's account growth and churn rates, see the following paragraphs for more details. 
Including also these measurements, we discovered a total of 3,546,479,731 accounts (i.e., distinct phone numbers).

\parvspace
\noindent
\textbf{Complementary Measurement: Growth and Churn.}
WhatsApp's user base is dynamic, with new accounts joining the platform and old ones being deleted.
To measure the extent of account growth and churn, we repeated the first step of \textit{account discovery} for four countries, namely Belgium (BE, 4x), India (IN, 3x), Iran (IR, 5x), and the US (3x).

\parvspace
\noindent
\textbf{Complementary Measurement: Profile Pictures.}
We relied on WhatsApp's XMPP protocol for data collection in our measurements.
However, this only provides the URL of a profile picture.
The actual picture needs to be fetched out-of-band via HTTP(S) from the domain \texttt{pps.whatsapp.net}, resolving into IP addresses of Meta's CDN.
We collected the picture URLs for all available accounts, but restricted the actual image downloads to phone numbers within the \texttt{+1} country code range.
This allows us to assess the general feasibility of crawling WhatsApp profile pictures at scale and analyzing the extent of available profile pictures.
For the download, we used 1,000 parallel threads on our server.
Testing for rate limits, we also experimented with more worker threads (up to 10,000), and WhatsApp's servers did not prevent us from collecting pictures.

\section{Results - Users and Devices}
\label{resultsusers}

In this section, we evaluate the dataset of 3.5\,B accounts collected in the first and second rounds of our measurements, i.e., accessing the endpoints \textit{GetDeviceList} and \textit{GetUserInfo}.
This includes the device indexes of main and companion devices, public profile picture, about tag line, related timestamps, and business profile information.
We refer to Section~\ref{sec:resultsprekeys} for an analysis that also includes the keying material used for E2EE.

\begin{table*}
\centering
{
\rowcolors{3}{lightgray!30}{}
\begin{tabular}{llrr||rrrrrrr}
\toprule
\multicolumn{4}{c||}{} & \multicolumn{6}{c}{(\%)} \\
 & Country & \# Accounts & Global Share & Android & iOS & Picture & About Text & Business & Companions \\

\midrule
1 & India & 749,075,246 & 21.67\,\% & 95 & 5 & 62.2 & 29.5 & 9.8 & 6.2 \\
2 & Indonesia & 235,245,077 & 6.81\,\% & 92 & 8 & 49.1 & 27.5 & 10.7 & 9.3 \\
3 & Brazil & 206,949,224 & 5.99\,\% & 81 & 19 & 61.1 & 41.5 & 10.3 & 15.5 \\
4 & United States & 137,859,284 & 3.99\,\% & 33 & 67 & 44.0 & 32.8 & 2.4 & 6.1 \\
5 & Russia & 132,855,022 & 3.84\,\% & 76 & 24 & 61.7 & 33.5 & 3.6 & 9.4 \\
6 & Mexico & 128,324,166 & 3.71\,\% & 82 & 18 & 46.1 & 23.3 & 4.1 & 11.7 \\
7 & Pakistan & 98,277,665 & 2.84\,\% & 95 & 5 & 58.5 & 20.0 & 21.7 & 5.4 \\
8 & Germany & 74,565,425 & 2.16\,\% & 58 & 42 & 51.0 & 35.4 & 2.2 & 13.4 \\
9 & Türkiye & 72,131,903 & 2.09\,\% & 73 & 27 & 48.0 & 33.4 & 3.0 & 12.0 \\
10 & Egypt & 69,317,806 & 2.01\,\% & 90 & 10 & 53.2 & 25.1 & 11.3 & 6.1 \\
11--245 & & 1,552,021,571 & 44.90\,\% & 77 & 23 & 56.9 & 27.9 & 9.3 & 9.0 \\
\midrule
\multicolumn{2}{l}{\textbf{GLOBAL (245 countries)}} & 3,456,622,389 & 100.00 & 81 & 19 & 56.7 & 29.3 & 9.0 & 8.8 \\
\bottomrule
\end{tabular}
}
\vspace{1ex}
\caption{Top 10 countries ranked by number of WhatsApp accounts. ¸\\ Android, iOS, picture, about text, business, and companions refer to their share in the respective country.}
\label{tab:result-overview-top-10}
\end{table*}

\parvspace
\noindent
\textbf{WhatsApp Accounts per Country.}
\label{sec:dataset}
Table~\ref{tab:result-overview-top-10} provides an overview of the collected data with detailed results for the ten countries with the most accounts.
A more verbose version is shown in Table~\ref{tab:result-overview-top-50} in the Appendix; the complete country overview can be inspected online\footnote{\url{https://github.com/sbaresearch/whatsapp-census}}.
India has the largest user population and accounts for 21\% of all WhatsApp accounts.
More than 57\,\% of world users have a public profile picture, with regional differences from 36\,\% in the Philippines to 80\% or more in some Western African countries (e.g., Mali, Burkina Faso, Guinea-Bissau).
Similarly, we were able to retrieve the about text from at least 29\,\% of global users, with numbers ranging from 6\,\% in Algeria and Philippines to more than 45\,\% in several European countries (e.g., the Netherlands, Sweden, Italy, Finland and UK).
The highest regional differences were discovered for the share of business accounts.
Globally, about 9\,\% are marked as business accounts.
While in some countries, this feature seems to be active for a considerable share of regular user accounts (e.g., roughly a third of all active accounts in various African countries), other regions show more authentic values (e.g., 2\,\% in the US or Germany).
A user can be marked as a business simply by setting up their account via the WhatsApp Business application\footnote{\url{https://play.google.com/store/apps/details?id=com.whatsapp.w4b}}, as there is no formal verification or entry barrier\footnote{All accounts on the business app are marked as business accounts; only verified ones receive an official checkmark.}.
Importantly, using the business app may impact a user's privacy, as the account's verified name, and optional attributes such as location (latitude/longitude), website, or description are always publicly visible.

\begin{figure}[t]
  \centering
  \definecolor{whatsapp-green-light}{HTML}{25D366}
  \definecolor{whatsapp-green-teal}{HTML}{075E54}
  \includesvg[width=\linewidth]{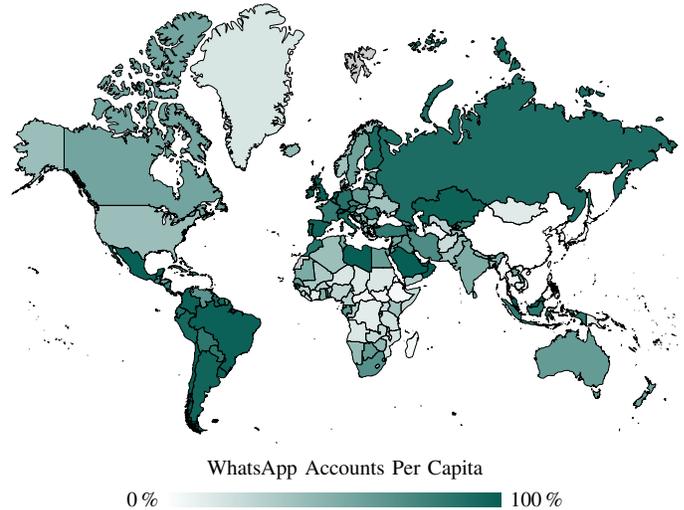}
  {\footnotesize WhatsApp Accounts Per Capita}\\
  {\footnotesize 0\,\%}~\tikz{\fill[left color=white,right color=whatsapp-green-teal] (0,0) rectangle (0.5\linewidth,0.5em);}~{\footnotesize 100\,\%}
  \caption{WhatsApp Use per Capita: At 95\,\% in South America and 80\,\% in Europe, a majority of citizens have an active WhatsApp account.}
  \label{fig:image-world-coverage-market-share}
  \vspace{-0.09cm} %
\end{figure}

\parvspace
\noindent
\textbf{WhatsApp Accounts per Capita.}
We calculated the number of active WhatsApp accounts per population for ISO~3166 countries using population data from the United Nations~\cite{UN_Population_WPP}, see Figure~\ref{fig:image-world-coverage-market-share}.
Monaco ranks first with an exceptional 480\,\%, i.e., 4.8 active WhatsApp accounts per resident. 
The United Arab Emirates follows with 176\,\%, aligning with market statistics~\cite{statista_uae_mobile_subscriptions} and media reports~\cite{aletihad_uae_digitalhabits}.
Our findings also reveal that WhatsApp is extremely popular in Latin America, with rates exceeding 98\% %
In contrast, we identified several countries with low market penetration --- Ethiopia (4\,\%), South Korea (4\,\%), and Japan (2\,\%).
The latter numbers are consistent with the presence of dominant local competitors --- \textit{KakaoTalk}~\cite{kakaoTalk_usage_korea} in South Korea and \textit{Line}~\cite{line_usage_japan} in Japan --- that are preferred over WhatsApp.
We provide further visualization per continent in
Figure~\ref{fig:market-penetration-continent-barchart}.

\begin{table}[tb]
\centering
\definecolor{highlight}{RGB}{210,250,210}
\newcommand\tH{\cellcolor{highlight}}
\begin{tabular}{@{}lrrr@{}}
\toprule
Date       & \# Users     & Per Capita & Companion Use \\ \midrule
2024-12-22 & 59,905,704   & 66.49\,\%          & 0.89\,\%          \\
2024-12-24 & \multicolumn{3}{c}{WhatsApp ban lifted \tH} \\
2025-01-06 & 62,330,344   & 69.18\,\%          & 1.73\,\%          \\
2025-01-20 & 63,829,328   & 70.84\,\%          & 2.15\,\%          \\
2025-02-16 & 65,600,472   & 72.81\,\%          & 2.46\,\%          \\
2025-03-19 & 67,114,098   & 74.49\,\%          & 2.55\,\%          \\ \bottomrule
\end{tabular}
\vspace{1ex}
\caption{Users in Iran (population 90.1\,M) despite WhatsApp Ban.}
\label{tab:iran-increase}
\end{table}

\parvspace
\noindent
\textbf{Countries Banning WhatsApp.}
As of December 2024, WhatsApp was officially banned in China, Iran, Myanmar, and North Korea.
Despite these restrictions, we found active accounts in these countries, for example, 2.3\,M Chinese phone numbers.
While this represents only a small fraction of China's total population of 1.4\,B, , media reports indicate that the use of WhatsApp in China may involve considerable personal risk~\cite{towey_2021_china, bbc_2021_china_video}.
Similarly, we found 1.6\,M accounts in Myanmar, equivalent to 3\,\% of its population.

In Iran, we identified 59\,M active accounts corresponding to two-thirds of its population (90.10\,M).
This substantial share aligns with previous statements from Will Cathcart~\cite{bbc_wwhatapp_bans_2024}, the current head of WhatsApp.
Coincidentally, the ban was officially lifted shortly after our first measurements~\cite{reuters_iran_lift_ban_2024}, 
and we repeatedly queried the country's number range to observe the development of WhatsApp in the country, see Table~\ref{tab:iran-increase} for our results.
In addition to a steady increase in registered phone numbers, the proportion of users with companion sessions (e.g., WhatsApp Web) tripled in the months after the ban was lifted.
The initially low share of companion devices may indicate that circumvention methods were primarily available on mobile devices, or that users avoided pairing desktop devices in public or work environments.

Lastly, a total of five active numbers were found within North Korea's number range.

\parvspace
\noindent
\textbf{About Text.}
While many accounts retain one of the default about tag lines provided by WhatsApp (e.g., \textit{``Hey there! I am using WhatsApp''}), others publicly reveal sensitive or personal information in their custom text.
For example, some users disclose their political ideology \textit{``Make America Great Again''}, indicate their sexual identity or orientation \textit{``LGBTQIA+''}, 
express religious beliefs by quoting verses from the Bible or Quran, or even reference drug use \textit{``Hey there! I am using cocaine''}.
We found hints of criminal activity, including drug dealers listing current offerings and prices, as well as an actual business account named \mbox{``SECRET TAXI''} advertising drugs with a detailed list of products \textit{``COKE-WEED-HASH-KETAMIN-MDMA-XTC-SPEED-LSD-2CB-3MMC-4MMC/MEMPHEDRON''}.

In other cases, users shared links to accounts on external platforms (e.g., LinkedIn, Instagram, Tinder, \mbox{OnlyFans}), which could be used to correlate and enrich profiles with additional meta-information.
Moreover, we found users posting their private or professional email addresses, some of them even being part of governmental or military organizations (e.g., \textit{@bund.de}, \textit{@state.gov}, \textit{@army.mil}, \textit{@navy.mil}, \textit{@us.af.mil}).
These practices significantly increase the risk of doxxing, identity exposure, and targeted attacks.

\parvspace
\noindent
\textbf{Profile Picture.}
Recent media reports revealed that high-ranking US government officials --- including Pete Hegseth (Secretary of Defense), Mike Waltz (National Security Adviser), and Tulsi Gabbard (Director of National Intelligence) --- were identified using their phone numbers as active on several messengers, including WhatsApp~\cite{beuth_2025_spiegel}.
According to the report, both Hegseth and Waltz had public profile pictures clearly showing their faces, which could be used to identify their accounts and correlate their phone number.

To demonstrate the scale and feasibility of this kind of data extraction from the platform, we downloaded 77\,M public profile pictures (overall 3.8\,TB of data) for all accounts within the American \texttt{+1} number range over a few hours.
We then applied an automated face detection algorithm using OpenCV to a randomized sample (N = 500,000), finding that roughly two-thirds of the images (66\,\%) contain detectable human faces.
In the hands of a malicious actor, this data could be used to construct a facial recognition--based lookup service --- effectively a ``reverse phone book'' --- where individuals and their related phone numbers and available metadata can be queried based on their face.
Beyond facial features, additional elements captured in profile pictures, such as license plates, street signs, or recognizable landmarks, could enable more sophisticated profiling and leak a user's identity, location, or daily environment.

\parvspace
\noindent
\textbf{Companion Usage.}
By leveraging the \textit{GetDeviceList} endpoint during \textit{account discovery}, we were able to automatically identify users who utilize companion sessions (e.g., WhatsApp Web).
Since companion devices are assigned auto-incrementing device indexes, this allows us to determine how many companion devices a user currently has and infer behavioral patterns. For instance, a stable set of device indexes (e.g., \verb|[0,1,3]|) suggests a consistent setup, whereas a sequence such as \verb|[0,79,86,93,95]| may indicate frequent use of ephemeral or short-lived sessions.
The prevalence of companion usage varies significantly across countries.
In some regions, such as the Netherlands or Israel, it is relatively common, with over 22\,\% of users having active companion sessions.
In contrast, it is insignificant in others, with usage rates below 3\,\% in countries like Algeria, Congo, Iraq, Iran, Syria, Tanzania, and Yemen.

Besides the simple adoption rate of companion devices, we can also compare the longevity and stability of the sessions.
For example, France and Indonesia both show a companion usage rate of around 10\,\%, yet their average device indexes (i.e., the current auto-increment values) are 8 and 31, respectively.
A similar pattern emerges when comparing Germany and Colombia, with comparable companion adoption (13--14\,\%), but substantial differences in average auto-increment values (DE: 12, CO: 34).
This suggests that Indonesian and Colombian users tend to create and discard companion sessions more frequently than their French and German counterparts.

\begin{table}[t]
\centering
\newcommand\tFa{$^{\ast}$}
\newcommand\tFb{$^{\mathrm{b}}$}
\newcommand{\roth}[1]{\rot{#1}}%

\begin{tabular}{@{}l|ccc@{}} \toprule
Client Implementation & Registration  & Signed PK   & One-Time PK  \\ \midrule

Android               & $R$           & 0           & $R$                       \\
iPhone                & $R$           & $R$         & 1             \\ \midrule
WhatsApp Web\tFa      & $R$ \& 0x3FFF & 1           & 1                 \\
DesktopApp macOS     & $R$           & $R$         & 1               \\
DesktopApp Windows   & $R$           & 1           & 1                 \\ \bottomrule
\end{tabular}
\\
\vspace{1ex}
$R$~Random number. \hspace{1ex} %
\tFa~Verified on Firefox, Chrome, Safari.\\
\vspace{1ex}
\caption{Init. values for key IDs as shown by Gegenhuber et al. \cite{gegenhuber_2025_prekeypogo}.}
\label{tab:characteristic-key-ids}
\end{table}

\parvspace
\noindent
\textbf{Account Net Growth and Churn.}
Most of our crawling was conducted as a single-shot measurement, capturing a snapshot of the registered user base.
We repeated our \textit{account discovery} measurements across multiple iterations for a small subset of selected target countries to gain insights into user dynamics over time.
Table~\ref{tab:user-churn} presents our longitudinal observations, reporting normalized net growth and user churn rates based on a 30-day average.

While all countries exhibited organic net user growth (ranging from 0\,\% to 1.5\,\% per month) across all longitudinal measurements, user churn (i.e., the share of users that leave the platform) varied significantly.
For instance, Belgium maintained a relatively stable user base, with losses of less than 1\,\% per month.
India demonstrated a higher churn rate of 2\,\% per month.
The highest churn rates were observed in the US, with peak losses exceeding 4\,\% of users in a single month.

\begin{table}[t]
\centering
\newcommand\tFa{$^{\dag}$}
\newcommand\tFb{$^{\ast}$}
\begin{tabular}{llrrrrr}
\toprule
     & \textbf{Crawl Date} & \textbf{Delta (days)} & \textbf{Net Growth\tFa} & \textbf{Churn\tFa} \\ \midrule
\multirow{5}{*}{\rotatebox{90}{Belgium}}
 & 2024-12-30 \\
 & 2025-01-20 & 21 & 0.28 & 0.71 \\
 & 2025-02-15 & 26 & 0.25 & 0.67 \\
 & 2025-03-19 & 32 & 0.22 & 0.64 \\
 & 2025-04-14 & 26 & 0.25 & 0.64 \\
\hline
\multirow{4}{*}{\rotatebox{90}{India}}
 & 2024-12-16 \\
 & 2025-02-16 & 62 & 0.36 & 1.80 \\
 & 2025-03-20 & 32 & 0.11 & 2.11 \\
 & 2025-04-16 & 27 & 0.63 & 1.62 \\
\hline
\multirow{6}{*}{\rotatebox{90}{Iran}}
 & 2024-12-22 \\
 & 2025-01-06 & 15 & 7.78 & 2.18 \\
 & 2025-01-20 & 14 & 5.03 & 1.75 \\
 & 2025-02-16 & 27 & 3.00 & 1.90 \\
 & 2025-03-19 & 31 & 2.18 & 1.70 \\
 & 2025-04-14 & 26 & 1.71 & 1.60 \\
\hline
\multirow{4}{*}{\rotatebox{90}{US}}
 & 2025-01-19 \\
 & 2025-02-17 & 29 & 0.41 & 4.25 \\
 & 2025-03-19 & 30 & 1.27 & 3.57 \\
 & 2025-04-14 & 26 & 0.54 & 4.34 \\
\bottomrule
\end{tabular}
\\
\vspace{1ex}
\tFa~Normalized to average percentage rates for 30 days.\\
\vspace{1ex}
\caption{Net growth and churn values for different target countries.}
\label{tab:user-churn}
\end{table}

\section{Results - Prekey Bundle Analysis}
\label{sec:resultsprekeys}

In this section, we evaluate the collected dataset of 
3.5\,B accounts, considering the third round of our measurements via the \textit{GetPrekeys} endpoints.
The so-called \emph{prekey bundle} is another data item retrievable for each device (both main and companion) regardless of any privacy settings.
Therefore, the section starts with an overview of the data points available in WhatsApp's prekey bundles and then provides various insights that can be drawn from the gathered data. %

\parvspace
\noindent
\textbf{WhatsApp Public Key Types.}
WhatsApp adopted the Signal protocol to provide E2EE~\cite{noauthor_whatsapp_2023}.
The Signal protocol is a suite of related protocols~\cite{marlinspike_x3dh_2016, perrin_double_2016, marlinspike_private_2014, marlinspike_sesame_2017, kret_pqxdh_2023} that have been the subject of extensive research across numerous studies~\cite{cohn-gordon_post-compromise_2016, alwen_double_2019, cohn-gordon_formal_2020, brendel_post-quantum_2022,wichelmann_help_2021, cremers_formal_2023, clone2020, gegenhuber_2025_prekeypogo}. For detailed information, we refer to the respective specifications and papers. 

For this work, it is sufficient to know that the handshake protocol (X3DH~\cite{marlinspike_x3dh_2016}) requires three types of X25519 public keys for each communicating device (main as well as companion), namely the long-lived \emph{identity key}, the medium-term \emph{signed prekey}, and the ephemeral \emph{one-time prekey}.
The identity key is not supposed to change over the device's lifetime, whereas a one-time prekey should be used only once, as the name suggests. 
The one-time prekey ensures perfect forward secrecy, also for the first messages when initiating a new conversation. 
Since instant messaging is asynchronous, recipients can be offline when a new communication session is initiated. 
Therefore, a stash of one-time prekeys is uploaded to the WhatsApp servers and replenished if needed. 
If no one-time prekeys are available, only identity and signed prekeys are used, which leads to imperfect forward secrecy of initial messages until the signed prekey is rotated. 
Therefore, these signed prekeys should be rotated periodically. 
Suggested intervals in the specifications range from a week to a month~\cite{noauthor_whatsapp_2023, marlinspike_x3dh_2016, meta2023messenger}. WhatsApp has chosen an interval of 30 days.

\begin{figure}[t]
  \centering
  \includegraphics[width=\linewidth,trim={0.1cm 0.25cm 0.1cm 0.1cm},clip]{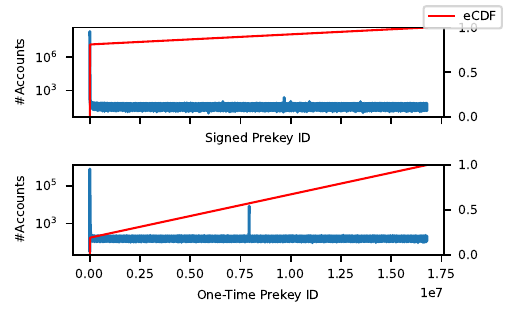}
  \caption{The OS-specific characteristic initialization values (Table~\ref{tab:characteristic-key-ids}) can be observed at scale.
  The empirical cumulative distribution functions (eCDFs) are left-skewed, reflecting the 0-based initialization patterns of the different IDs which are typical for the operating systems Android (0-initialized signed prekey ID) and iOS (0-initialized one-time prekey ID).
  The eCDF in the top figure corresponds to Android devices, which account for 81\,\% of the observed user base, while the bottom figure represents iOS devices, comprising the remaining 19\,\%. The spike in this figure is investigated in Section~\ref{sec:prekey_bundle_analysis}.}
  \label{fig:key-ids-distribution}
\end{figure}

\parvspace
\noindent
\textbf{OS Market Statistics.}
Previous work~\cite{gegenhuber_2025_prekeypogo} identified distinct initialization patterns in prekey IDs across different client implementations of WhatsApp, as summarized in Table~\ref{tab:characteristic-key-ids}.
Building on this, we first verified these OS-specific patterns using a controlled set of test devices.
We then leveraged the signed and one-time prekey IDs of bundles collected from the discovered accounts to estimate the global OS market share among WhatsApp users at scale.

Figure~\ref{fig:key-ids-distribution} presents the distributions of \textit{signed} and \textit{one-time prekey IDs} across all retrieved accounts, demonstrating that OS classification is %
feasible via both ID types.
Our analysis across 3.5\,B accounts reveals an Android share of 81\,\% and iOS at 19\,\%, as illustrated in the eCDFs in Figure~\ref{fig:key-ids-distribution}. 
The numbers vary per continent, see Figure~\ref{fig:sunburst}, and indicate higher prevalence for iOS in regions with higher purchasing power.
The information on the OS is available per phone number and could be abused by malicious actors for reconnaissance or targeted exploits for a specific OS.
Apple users are often considered higher-value targets --- both for commercial profiling and in the context of scams or ransomware --- due to their alleged purchasing power.

\begin{figure}
    \centering
    \includegraphics[width=\linewidth,trim={0.2cm 0.25cm 0.2cm 0.2cm},clip]{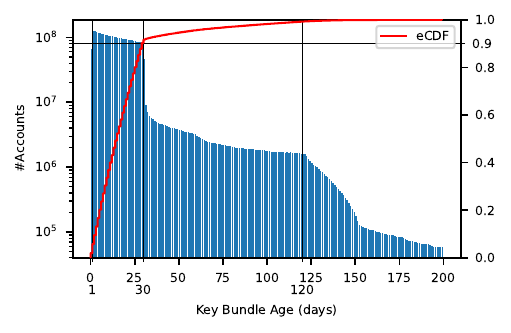}
    \caption{Distribution of prekey bundle age across our retrieved data. Over 90\,\% of users have updated their keys within the past month. Consistent with WhatsApp's account deletion policy, most accounts with keys older than 120 days have been automatically removed.}
    \label{fig:account-activeness}
\end{figure}

\parvspace
\noindent
\textbf{Verifying Account Activeness.}
We want to ensure that the retrieved phone numbers correspond with WhatsApp accounts that are actively used, rather than outdated or inactive accounts.
Every prekey bundle retrieved via the \textit{GetPrekeys} endpoint also contains an epoch timestamp updated every time the corresponding client device pushes new keys to the server.
The update occurs when the user updates i) their identity key (e.g., when migrating the account to a new smartphone), ii) their signed-prekey (automatically done once a month), or iii) their one-time prekey batch (necessary when the previous batch is depleted).
Since the main device is expected to be consistently online, the timestamp of a prekey bundle for an active account should be younger than one month.
Figure~\ref{fig:account-activeness} shows a histogram of the bundle age of the retrieved main device prekeys.
As anticipated, more than 90\,\% of accounts have a recent prekey bundle that was updated within the last month.
Moreover, there is a noticeable dip in the histogram at 120 days, corresponding to WhatsApp's deletion policy for inactive accounts\footnote{\url{https://faq.whatsapp.com/828406668498455}}.
Although only representing an insignificant share in the overall distribution, some accounts appear to be not automatically removed after 120 days.

\begin{figure}
    \centering
    \includegraphics[width=\linewidth,trim={0.2cm 0.25cm 0.2cm 0.2cm},clip]{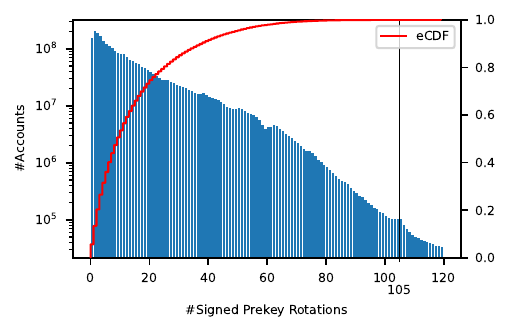}
    \caption{On Android devices, auto-incrementing signed prekey IDs can be used to estimate device age.}
    \label{fig:device-age}
\end{figure}

\parvspace
\noindent
\textbf{Estimating Device Age and Popularity.}
WhatsApp rotates each device's signed prekey once per month, incrementing the signed prekey ID by one with each update.
On Android, this ID starts at 0, allowing the approximation of a device's age in months by examining its current value.
Figure~\ref{fig:device-age} presents a histogram of signed prekey IDs of Android devices across our dataset.
Most devices appear to be relatively recent (roughly 80\,\% have been created within the last two years).
A distinct bump occurs at 105 months. Subtracting this from the date of our measurements (February 2025) points to April 2016, the month in which WhatsApp introduced E2EE to their platform~\cite{metz_2016_whatsapp_encryption}.
We can only speculate about values above the 105-month marker.
They either represent iOS devices (randomly choosing their signed prekey ID) or Android devices that have undergone atypical key rotations, potentially triggered by incorrect system clocks or security-related reinitialization routines.

Similarly, by analyzing the one-time prekey IDs on iOS devices, we can estimate how frequently they uploaded new prekey batches to the WhatsApp server.
Since a prekey is consumed whenever a device initiates a conversation with a new contact, their depletion rate offers a valid approximation of a device's popularity.
Figure~\ref{fig:device-chattiness} in the Appendix illustrates the observed distribution across our dataset.
Knowing the device age and popularity could be exploited to identify long-term or highly active users --- individuals who may hold higher value from a marketing or targeting standpoint.

\begin{figure}[t]
    \centering
    \includegraphics[width=0.98\linewidth,trim={0.2cm 0.25cm 0.2cm 0.2cm},clip]{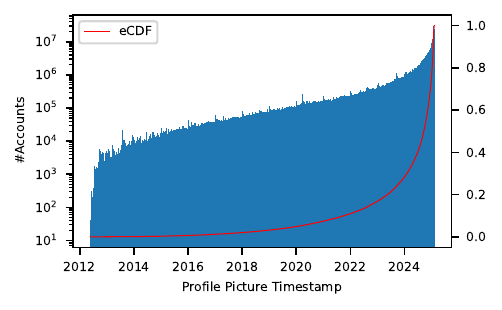}
    \caption{Distribution of profile picture timestamps. Most users have recently updated their picture.}
    \label{fig:timestamp-picture}
\end{figure}

\begin{figure}[t]
    \centering
    \includegraphics[width=0.98\linewidth,trim={0.2cm 0.25cm 0.2cm 0.2cm},clip]{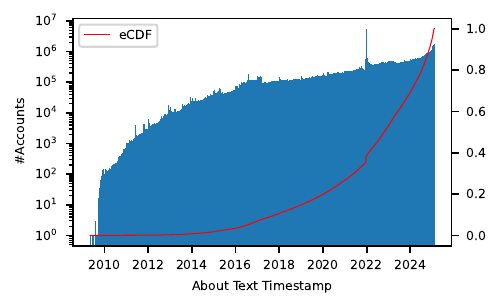}
    \caption{Distribution of about status update timestamps. About messages are updated less frequently on average compared to profile pictures.}
    \label{fig:timestamp-status}
\end{figure}

\parvspace
\noindent
\textbf{Identifying Longterm Users.}
Besides estimating a user's device age using the prekey bundle, we can also investigate the overall account age by inspecting the timestamps when a user updated their profile picture or about text.
Figures~\ref{fig:timestamp-picture} and~\ref{fig:timestamp-status} illustrate the distribution of these timestamp values.
These timestamps are upper and lower bounds for a user's earliest and latest activity.
Our large-scale data indicates that users update their profile pictures more frequently than their about messages.
The noticeable spike in about status updates around November 2021 likely corresponds to WhatsApp's revised privacy policy\footnote{\url{https://www.whatsapp.com/legal/privacy-policy-eea/revisions/20211122}}, which followed a significant GDPR fine and received broad media attention~\cite{bbc_2021_privacy_policy} as EU and UK users were notified about the change directly in the app.

\section{Results - Public Keys Analysis}
\label{sec:prekey_bundle_analysis}
While the prekey bundle metadata consisting of timestamps and public key IDs was already analyzed in Section~\ref{sec:dataset}, this section focuses on the actual X25519 public keys and their signatures required for the X3DH handshake protocol~\cite{marlinspike_x3dh_2016,noauthor_whatsapp_2023} which are also included in a prekey bundle: 
Namely the long-term \emph{identity key}, the medium-term \emph{signed prekey}, and an ephemeral \emph{one-time prekey}. 
As a first step, we verified all Ed25519 signatures on the signed prekeys within these bundles to validate the integrity of the retrieved data. 
We encountered only two invalid signatures across the entire dataset, which supports a high level of confidence in the integrity of the collected data.

\parvspace
\noindent
\textbf{Missing One-time Prekeys.} 
As mentioned in Section~\ref{sec:resultsprekeys}, the absence of a \textit{one-time prekey} weakens perfect forward secrecy (PFS) for the initial messages in a new conversation.
Therefore, we assessed the prevalence of prekey bundles without \textit{one-time prekey} in our dataset.
We found that 0.8\,\% of main devices and 13\,\% of companion devices did not have a \textit{one-time prekey} at the scan time. 
One possible interpretation of this result is that companion devices, such as desktop computers, are more likely to be left turned off for a while. During this time, they are not able to replenish consumed one-time prekeys on WhatsApp servers. While main devices are smart phones that are usually always online. 
This finding illustrates that one-time prekey exhaustion is indeed common for companion devices in practice, even without targeted depletion attacks as described in~\cite{gegenhuber_2025_prekeypogo}.

\begin{table*}
\centering
{
\rowcolors{3}{lightgray!30}{}
\begin{adjustbox}{width=1\textwidth}
\begin{tabular}{crrrrrrl}
\toprule
\textbf{Pubkey} & \textbf{Occurances} & \textbf{Companion} & \textbf{IDkey} & \textbf{SIGprekey} & \textbf{OTprekey} & \textbf{\# Countries} & \textbf{Countries (Top 4)}\\
\midrule
\texttt{7aec...242d} & 131,406 & 8,682 & 0 & 115,784 & 15,622 & 141 & MM:38,207; ID:25,997; NG:13,483; PK:8,473 \\
\texttt{0df7...2d09} & 83,434 & 83,434 & 29,048 & 29,050 & 25,336 & 63 & NG:73,665; PL:3,684; ID:1,874; PK:1,002 \\
\texttt{d513...3565} & 4,080 & 0 & 4,080 & 0 & 0 & 1 & ID:4,080 \\
\texttt{ac30...b22f} & 1,460 & 0 & 1,460 & 0 & 0 & 1 & ID:1,460 \\
\texttt{667e...0876} & 1,295 & 0 & 0 & 1,295 & 0 & 1 & ID:1,295 \\
\texttt{6b7d...1e17} & 949 & 0 & 949 & 0 & 0 & 1 & ID:949 \\
\texttt{32e2...270a} & 781 & 0 & 0 & 781 & 0 & 1 & ID:781 \\
\texttt{13bc...0b7f} & 722 & 0 & 722 & 0 & 0 & 4 & IR:719; RO:1; AM:1; AU:1 \\
\texttt{d667...0b35} & 688 & 0 & 688 & 0 & 0 & 1 & ID:688 \\
\texttt{e03a...777e} & 581 & 18 & 581 & 0 & 0 & 10 & SA:559; EG:9; YE:3; SD:3 \\
\midrule
\textbf{TOTAL} & 2,957,433 & 105,143 & 1,978,340 & 935,051 & 44,042 & 225 & (distinct)\\
\bottomrule
\end{tabular}
\end{adjustbox}
}
\vspace{1ex}
\caption{Top 10 of reoccurring public keys.  }
\label{tab:top_collisions}
\end{table*}

\parvspace
\noindent
\textbf{Key Reusage.} 
Ideally, no public key occurs in more than one device at the same time.
Recurring public keys may indicate errors in the distribution of (one-time) prekeys, faulty implementations or configurations~\cite{gegenhuber_2024_diffie}, or broken random number generators. 
To search for duplicate public keys, we generated a Bloom filter~\cite{bbloom_1970} containing all public keys to compare each public key against all other public keys in constant time and logged whether a public key was flagged as already being part of the filter. 
We over-provisioned the Bloom filter to minimize the theoretical possibility of false positives (false positive rate of $1e-12$). 
Moreover, any public key flagged as being part of the filter at least twice in the log file can definitely be considered as re-used. 

Using this method, we discovered %
2.3\,M distinct X25519 public keys that have been re-used in prekey bundles of
over %
2.9\,M distinct devices.
Table~\ref{tab:top_collisions} lists the public keys with the highest number of occurrences. As shown in the table --- and further illustrated in Figure~\ref{fig:country_code_hist_7ac28} 
--- the most frequently reused public keys are predominantly associated with phone numbers from Myanmar and Nigeria. 
This indicates either a problem with a (custom) WhatsApp client used in these regions, or may point to fraudulent activity involving large-scale creation of accounts tied to different phone numbers.

Reusing public keys implies that the corresponding secret keys are also reused, which is particularly concerning when the devices involved belong to different entities.
A compromised identity key effectively results in full account compromise with respect to the confidentiality of conversations. It enables an attacker to cryptographically sign updates to the device list and authorize the addition of new companion devices, thereby gaining the ability to intercept or participate in encrypted communications.
We even identified cases in which all three public keys in a prekey bundle —namely, the identity key, signed prekey, and one-time prekey—  collided with those of other devices, further highlighting the severity of key reuse.
Interestingly, key reuse is not evenly distributed across the different key types.
The \textit{identity key} is by far the most affected, with %
2\,M instances of reuse.
This uneven distribution suggests that the majority of key reuses are unlikely to be caused by faulty random number generators --- otherwise, we would expect similar reuse rates across all three key types.

\parvspace
\noindent
\textbf{Number Change.} 
Despite specific keys being reused in high numbers, 97\,\% of distinct public keys have only been reused once or twice (see Figure~\ref{fig:hist_coll_count_100} in the appendix). 
If only main devices are considered, 68\,\% of reuses affect identity keys, which suggests a usage pattern during regular operation as a root cause.
Consequently, we looked closer at identity keys reused exactly twice in main devices.
From these identity public keys re-uses, 96\,\% are attributed to two phone numbers within the same country, suggesting phone number change is a plausible explanation for the majority of cases.
While this behavior does not constitute a security vulnerability per se, it can be a potential privacy issue, as the reuse of public keys enables linkage of users across phone number changes.

To investigate legitimate root causes for duplicate identity keys, we monitored the prekey bundle of our device while migrating a WhatsApp account to another phone number\footnote{This is supported via a built-in dialog in the official WhatsApp client.}.
We observed that both the \textit{identity} and \textit{signed prekey} remained unchanged throughout the migration.
Additionally, we were able to reproduce scenarios in which previously installed keys --- left behind after logging out of WhatsApp --- were reused when setting up a new account with a different phone number on the same device.
This behavior suggests poor practices in account initialization.

\parvspace
\noindent
\textbf{One-time Prekey Reuse.}
While the previous paragraph offers a plausible explanation for the reuse of identity keys, there is no reasonable explanation for the reuse of \emph{one-time prekeys}, which, by definition, should never be reused --- neither on the same device nor different devices.
Their recurrence indicates a failure in the respective implementations.
The top two entries in Table~\ref{tab:top_collisions} show public keys which also re-occur as one-time prekeys. 
Interestingly, public key number two (\texttt{0df7...}) only occurs in companion devices, but here as each type of public key. 
A possible explanation for this would be a faulty non-official implementation of the WhatsApp companion protocol. 
Apart from these keys, we also discovered 2303 distinct one-time prekeys that are reused 3082 times.
Manual inspection of some of these cases indicates that the respective phone numbers belong to the same entity, as they originate from the same country and share identical about tag lines or business names. %
Moreover, the ID value of the reused \textit{one-time prekeys} also matches across different phone numbers.
This suggests that there is indeed a connection between the affected devices, while the root cause is still unclear.

\parvspace
\noindent
\textbf{All Zero Private Key.} 
Especially interesting are public keys which occur as different key types e.g., a identity key which also occures as signed prekey. 
There are three public keys in total, which (re-)occur in different key types. 
Key number two from Table~\ref{tab:top_collisions} (\texttt{0df7...}), a key recurring only once as signed prekey and one-time prekey for the same device (\texttt{ce8d...}), 
and a third one re-occurring 42 times in 20 distinct phone numbers, mainly from the US\footnote{Public key in raw hex:\\ 2fe57da347cd62431528daac5fbb290730fff684afc4cfc2ed90995f58cb3b74}.
For this public key, we were able to recover the corresponding private key, which consists of all bytes set to zero.
This, as well as the fact that for some of the affected devices this key occurred as identity key, signed prekey, and one-time prekey at the same time, suggests that the random number generation of the used implementation was broken.
To test for more subtle failures regarding random number generation, we generated private keys and computed the corresponding public keys to compare them against our Bloom filter. 
Like Bos et al.~\cite{bos_elliptic_2014} we tried all private keys with a Hamming-weight of up to $ 3 $, i.e., $\binom{256}{3}$ scalars, without finding any further matches between the self-generated public keys and the collected ones.

\section{Relation to the 2021 Facebook Data Leak}
\label{sec:comparison-facebook-leak}

\begin{table}[t]
\centering
\begin{tabular}{@{}lrrr@{}}
\toprule
Country & Active Numbers     & Active + in FB Leak      & Share \\ \midrule
EG      & 69.3\,M            & 25.9\,M             & 37.4\,\%    \\
IT      & 55.6\,M            & 21.6\,M             & 38.9\,\%    \\
US      & 137.9\,M           & 19.2\,M             & 13.9\,\%    \\
SA      & 38.9\,M            & 16.3\,M             & 41.9\,\%    \\
FR      & 54.0\,M            & 14.2\,M             & 26.2\,\%    \\ \bottomrule
\end{tabular}
\vspace{1ex}
\caption{Top Countries with still active numbers from Facebook Leak.
}
\label{tab:facebook-leak-comparisn}
\end{table}

Containment efforts following public data leaks are considered difficult.
In practice, once-leaked data remains accessible online, e.g., on specialized web forums or decentralized torrent networks.
In this section, we compare data from a six-year-old data leak with the collected WhatsApp data to investigate whether once-leaked data remains valid over extended periods and is a reliable basis for spam, phishing, or robocall attacks.

\parvspace
\noindent
\textbf{The Facebook Data Leak.}
In April 2021, a dataset containing the personal information of over 500\,M Facebook users from more than 100 countries was publicly leaked~\cite{holmes_2021_facebook_leak}, making it one of the largest scraping incidents to date.
According to media reports and an official statement from Meta~\cite{facebook_leak_2021}, the data was obtained through scraping activities conducted on the platform in 2019\footnote{The IDPC categorized this leak as scraping incident and imposed a 265\,M EUR fine for failing to implement adequate technical and organizational safeguards~\cite{dpc_2022_facebook}.}.
The dataset consists of 105 archives.
Each archive represents a country and contains one or more CSV files lacking a consistent format, character encoding, or character escaping, making parsing a challenge.
Our analysis shows that basic attributes, such as phone number, Facebook user ID, gender, first and last name, are present in almost all entries.
Beyond that, most entries contain additional information, such as relationship status, employer, email, birthday, birthplace, education, and residence.

\parvspace
\noindent
\textbf{Method for Comparison.}
We extracted all phone numbers from the dataset and validated them using \textit{libphonenumber}. 
This resulted in 488,410,549 unique phone numbers, with 32\,M originating from the US.
We checked whether these phone numbers were available in our WhatsApp dataset.
If so, we assumed the phone number was still active and therefore worthwhile for attackers.

\parvspace
\noindent
\textbf{Active Phone Numbers.}
We were able to find 58\,\% (281/488\,M) of the phone numbers exposed in the previous Facebook leak in our dataset, i.e., they are still active.
This highlights the long-term impact of data leaks, even more than six years after the original scraping.
Table~\ref{tab:facebook-leak-comparisn} lists the top five countries with the highest number of still active, leaked phone numbers.
In Egypt, Italy, and Saudi Arabia, around 40\% of currently active WhatsApp numbers also appear in the leaked dataset.
Inclusion in such a leak significantly increases the chance of being targeted by robocalls, marketing campaigns, or scams.
With phone number porting widely available, users may keep the same number for years, potentially enabling targeted phishing or social engineering using leaked data such as names, birthdates, emails, or employers.

\section{Related Work}
\label{sec:relatedwork}

\parvspace
\noindent
\textbf{Enumeration Attacks.}
In enumeration attacks, a search space of phone numbers or email addresses is (entirely) enumerated to find active accounts.
They have been applied to online social networks and messaging services at least since 2010 and are, as our work emphasizes, still applicable.
In 2010, Balduzzi et al.~\cite{Balduzzi2010Abusing} fed 10.4\,M email addresses in the find-friends functionality of eight social networks, namely Facebook, MySpace, Twitter, LinkedIn, Friendster, Badoo, Netlog, and XING.
The input email addresses were taken from a previous data breach.
Additional email addresses were created from a user's friends, e.g., \url{firstname.lastname@gmail.com}.
Three platforms then provide phone numbers as an output, besides other information.
Schrittwieser et al.~\cite{schrittwieser_2012_guess} were the first to enumerate phone numbers to find WhatsApp accounts in 2012.
They decided on a US range of 10\,M numbers and found 21,095 active accounts in 2.5 hours.
Cheng et al.~\cite{Cheng2013Bind} enumerated two Chinese number ranges on WeChat. From the 100,000 numbers, 14,179 were found to be active.
Kim et al.~\cite{Kim2015Design} probed KakaoTalk, a Korean messenger, and found 50,680 from a total of 101,000 numbers to be active.
Gupta et al.~\cite{Gupta2016Emerging} uploaded more than 1\,M Indian phone numbers on TrueCaller and Facebook, and found more detailed information for 722,696. A total of 51,409 were also found to be active on WhatsApp.
Kim et al.~\cite{Kim2017Hello} inserted phone numbers in Facebook's search field and look whether a profile is returned.
The authors investigated phone ranges, one in the US and another in Korea, and found 82,082 active from 430,384.
The most recent work dates from 2021:
Hagen et al.~\cite{hagen_2021_all} enumerated 50.5\,M addresses and found 5\,M active.
As an input set, the authors decided on US ranges of cellar operators and randomly sampled this address space.

Previous work is insofar similar as they exploited intended modes of interaction with the investigated platform and inserted the candidates via the smartphone's address book~\cite{schrittwieser_2012_guess,hagen_2021_all}, the app's export function~\cite{Kim2015Design}, the platform's search function~\cite{Kim2017Hello}, or a Gmail account~\cite{Balduzzi2010Abusing}.
In multiple cases, interaction with the app is automated~\cite{Kim2017Hello,hagen_2021_all}.
In comparison, we access WhatsApp via a reverse-engineered API, enabling fast enumeration of 63\,B candidates.
Additionally, we are able to directly investigate WhatsApp's protocol and infrastructure, revealing more detailed insights into its user population.

Table~\ref{tab:enumerationattacks} compares the enumeration attacks regarding the total search space, active addresses, total time of measurement, and speed in probes per second.
The total time is provided in varying granularity.
Yet, it allows an approximation of the measurement speed, i.e., how many numbers were tested for activity per second.
Our work outperforms previous studies regarding comprehensiveness --- the measurement covers a search space of 63\,B phone numbers with practically global coverage --- and speed.

\begin{table}[tb]
    \centering

    \newcommand\tFa{$^{\dag}$}
    \begin{tabular}{@{}rrrrrr@{}}
    \toprule
      Work & Year & \# Candidates  & Hits & Duration & Speed\tFa \\ \midrule  %
      \cite{Balduzzi2010Abusing} & 2010 & 10,400,000 & 876,000 & 2w & 9 \\ %
      \cite{schrittwieser_2012_guess}  & 2012   & 10,000,000  & 21,095 & 2.5h & 1,100 \\ %
      \cite{Cheng2013Bind} & 2013 & 100,000 & 14,179 & - & - \\ %
      \cite{Kim2015Design} & 2015 & 101,000 & 50,670 & - & 2 \\ %
      \cite{Gupta2016Emerging} & 2016 & 1,162,696 & 722,696 & - & - \\ %
      \cite{Kim2017Hello} & 2017 & 430,384 & 82,082 & 15d & $<$ 1  \\ %
      \cite{hagen_2021_all} & 2021 & 50,500,000 & 5,000,000 & 34d & 18 \\ \midrule %
      Our & 2025 & 63,170,000,000 & 3,546,479,731 & 27d & 7,000 \\ \bottomrule
    \end{tabular}
    \\
    \vspace{1ex}
    \tFa~Accounts per second. \\
    \vspace{1ex}
    \caption{Comparison of Enumeration Attacks. Our work outperforms previous approaches in terms of comprehensiveness and speed.}
    \label{tab:enumerationattacks}
\end{table}

\parvspace
\noindent
\textbf{Privacy Leaks Despite Message Encryption.}
Messaging services increasingly rely on encryption, even end-to-end encryption.
Nevertheless, multiple privacy leaks have been discovered.
Coull and Dyer~\cite{Coull2014Traffic} inferred user actions, message language, and length on iMessages from simply analysing the size of encrypted packets,
Park and Kim~\cite{Park2016Encryption} did similarly on KakaoTalk, and Bahramali~\cite{Bahramali2020Practical} identified administrators and members of dedicated channels on multiple services.
Schnitzler et al.\cite{schnitzler_hope_2023} exploit receipt messages to extract a user's geographic position from the observed round-trip times.
Gegenhuber et al. demonstrated that a subscriber's country-level location can be inferred from variations in ringback-tone frequencies in the traditional cellular layer~\cite{gegenhuber_2023_mobileatlas}.

With the advent of E2EE, a sender has to encrypt a message separately for each receiver device.
The messaging infrastructure provides the public keys for each device.
Be'ery~\cite{beery_whatsapp_2024} found this a privacy leak revealing an account's device setup information to anybody, even those not in the contact list or blocked.
Likewise, all receiver devices return individual reading receipts.
Their timing reveals whether a device is currently on or off.
Gegenhuber et al.~\cite{gegenhuber_2024_carelesswhisper, gegenhuber_2025_ccs} found a way to trigger stealth receipts, i.e., without notifying the user, and created activity profiles of the victim.
Used excessively, their action could even drain a victim's battery.
There is also the possibility of exhausting prekey material, degrading forward secrecy for new connections~\cite {gegenhuber_2025_prekeypogo}.

\section{Discussion}
\label{sec:discussion}
\noindent

\noindent
\textbf{Centralization and Monopoly in Messaging Apps.}
The utility of a messaging system for day-to-day communication relies on users being able to reach their peers. 
This dynamic naturally drives centralization and fosters monopolies, making it hard for new players to enter the market\footnote{The EU classified WhatsApp as a \emph{gatekeeper} according to the Digital Markets Act (DMA)~\cite{EU_Digital_Markets_Act_2022}, making such companies subject to new obligations, such as interoperability. Unfortunately real-world adoption remains limited as of today and not all features are available yet, e.g., group chats and voice/video calls~\cite{Meta_2024_DMA_interoperability,Hodgson_2024_Matrix_WhatsApp_DMA}.}.
Due to its current position, WhatsApp inherits a responsibility akin to that of a public telecommunication infrastructure or Internet standard (e.g., email). 
However, in contrast to core Internet protocols which are governed by openly published RFCs and maintained through collaborative standards --- this platform does not offer the same level of transparency or verifiability to facilitate third-party scrutiny.
This, as well as current requirements within the EU for interoperability~\cite{Meta_2024_DMA_interoperability}, underscore the need 
to uncover silent flaws that could pose risks to all users of the platform as well as to users of interlinked services.

\parvspace
\noindent
\textbf{Data Leaks.}
Our results prove the long-lasting and irreversible consequences of data leaks. Remarkably, more than half of the phone numbers previously leaked remained active even six years after being extracted from the platform.
Large platforms that handle the personal data of millions have a significant responsibility to prioritize data protection, thereby justifying the need for robust data protection legislation.
Leaked data may increase the risk of users being targeted by scammers or marketing campaigns.
 In more severe cases, the exposure of a user's presence on a platform could result in serious repercussions, particularly in countries where the service is banned by government authorities.

\parvspace
\noindent
\textbf{Far-reaching Survey Possibilities.}
Despite collecting only a limited set of data attributes per number, we were able to derive meaningful insights—both at a macro level and for individual users.
A user's OS information could be exploited as part of a sophisticated attack chain (e.g., for reconnaissance), while other attributes may leak details such as interests (about text), perceived age, gender, or race (profile picture), or geographic location (e.g., US state inferred from the phone number).
This type of data could be leveraged for highly targeted marketing or political propaganda, as seen in the Cambridge Analytica scandal~\cite{guardian_2018_facebook_cambridge_analytica}.

\parvspace
\noindent
\textbf{Public Key Collisions.}
We identified both weak (all-zero) and duplicate private keys being used across multiple devices on the messaging platform. The frequency of some of these key reoccurrences --- in certain cases involving over a hundred thousand distinct phone numbers --- is highly improbable under normal operational circumstances.
For many of these collisions, we observed indicators of fraudulent activity (e.g., romance scams or cryptocurrency fraud), which are known issues previously reported in the affected regions~\cite{aljazeera_2024_myanmar_scammills, nyt2025_scam_centers_asia, cnn_2025_myanmar_crackdown}. These anomalies could have been --- and arguably should have been --- detected more effectively by the platform operator.

\subsection{Limitations and Future Work}
In this study, we measured real-time systems that are continuously evolving, such as users joining or leaving the platform, gradually changing client implementations, and phone numbers being ported across different operators.
We utilized open-source tools (e.g., \textit{libphonenumber} and \textit{whatsmeow}) to obtain dynamic snapshots throughout our measurement campaign.
While we are confident that our dataset is sufficiently large to support reliable macroscopic insights, some blind spots remain (e.g., when users left the platform between our \textit{account discovery} and \textit{metadata retrieval} scans).

While WhatsApp is the most widely used instant messaging application globally, it is not the only one.
Exploring enumeration vulnerabilities on other messaging platforms --- especially as they adopt E2EE messaging --- would be a valuable direction for future research.
Such an investigation could reveal whether these platforms implement stricter rate-limiting mechanisms or if they are similarly susceptible to large-scale user discovery.
Like WhatsApp, many messaging services are developed and operated by commercial entities.
Even when parts of their protocols are publicly documented or formally verified, the real-world implementations and deployments often remain opaque and potentially contain security issues.
In this context, large-scale empirical studies serve as an important auditing mechanism, offering independent validation and the potential to uncover issues such as weak or poorly randomized key material.

\subsection{Countermeasures.}
At the individual user level, there are only two possible countermeasures:
(i) configure more restrictive privacy settings (e.g., non-public picture and about text) in WhatsApp,
(ii) deactivate WhatsApp and switch to an interoperable instant messenger~\cite{Meta_2025_messaging_interoperability,Hodgson_2024_Matrix_WhatsApp_DMA, EU_Digital_Markets_Act_2022} to still interact with peers on WhatsApp. To the best of our knowledge, the latter option is theoretically only available from within the EU and currently restricted to a limited feature set~\cite{Meta_2024_DMA_interoperability}. 

At an operator level the following countermeasures could be considered:

\parvspace
\noindent
\textbf{Rate Limiting.}
As previous work~\cite{hagen_2021_all} already showed, rate-limiting is a powerful way to mitigate large-scale account enumeration in instant messengers. 
Special care must be taken to avoid over-blocking of legitimate contact discovery attempts. 
As in our example, enumerating entire populations of foreign countries seems highly unlikely in typical day-to-day use cases originating from individual clients and could therefore be considered detectable.

\parvspace
\noindent
\textbf{Private Contact Discovery.}
Since contact discovery is needed for legitimate users to find and message new peers, it is generally hard to completely eliminate account enumeration possibilities on instant messaging platforms.
Private Set Intersection (PSI) protocols~\cite{DBLP:conf/eurocrypt/FreedmanNP04} enable the computation of the intersection between the registered user database and a users’ address books in a privacy-preserving manner, not leaking any phone numbers to the operator.
However, even with PSI, enumeration attacks remain an issue. Actively secure constructions --- those that preserve privacy even in the face of arbitrary deviations from the protocol --- do not prevent a party from using different inputs in each execution. Therefore, PSI protocols must be complemented with safeguards against enumeration attacks, such as limiting the number of protocol executions and minimizing the size of input sets~\cite{hagen_2021_all}.
Although, privacy-preserving approaches for contact discovery have been considered, for example based on PSI~\cite{kales_mobile_2019} or trusted hardware~\cite{signal2017private}, their deployment is non-trivial given necessary set sizes and practical constraints~\cite{hagen_2021_all}.

\parvspace
\noindent
\textbf{Encrypting Profile Information.}
The Signal messenger protects user-associated profile data using a dedicated \textit{profile key}~\cite{chase_signal_2020} to encrypt profile information (e.g., avatar images and profile names). This key is generated by the data owner and is only shared with contacts who have received a message from that person. As a result, even the service operator cannot access this information, effectively preventing the exposure of profile data, even in the event of a large-scale data leak.
In addition to encrypting profile data, WhatsApp could adopt a \textit{privacy by default} approach by hiding such information unless users explicitly choose to share it through their privacy settings\footnote{According to WhatsApp, the default setting for the about tagline has been \textit{Contacts Only} for new registrations since 2023.}.

\parvspace
\noindent
\textbf{Harmonizing Client Implementations.}
We were able to exploit differences in WhatsApp's Android and iOS implementations to infer a user's operating system.
Harmonizing the underlying codebases for the different operating systems would mitigate such information disclosure side-channels.

At a high level infrastructure perspective, the general design and length of phone numbers could be reconsidered:

\parvspace
\noindent
\textbf{Phone Number Space.}
Theoretically, an attacker’s ability to successfully enumerate users' phone numbers could be reduced by increasing the plausible number space within a country --- similar to the challenges of discovering IPv6 addresses compared to IPv4~\cite{DBLP:conf/woot/UllrichKHDW14}. For instance, Austria, with a population of around 9\,M, has over 500\,B possible phone numbers. Thus, scanning Austria's entire number space would have taken us nearly a year. However, we were able to identify used number blocks in Austria based on known issued numbers, which highlights the need for random assignment of phone numbers across the entire range. Considering that it is a legacy infrastructure, significant improvements and effective structural changes to the entire phone number system are unlikely to manifest in the near future. Moreover, given the persistent nature of phone numbers and their (necessary) propagation, such measures would have no impact on already leaked numbers and thus pose no immediate benefit for a large part of the population.

\section{Conclusion}
\label{sec:conclusion}
In this paper, we developed a method to generate plausible mobile phone numbers for 245 countries and validated the feasibility of large-scale phone number enumeration for the popular mobile messenger application WhatsApp.
This way, we could enumerate 3.5\,B active user accounts, including phone numbers, timestamps, public keys and signatures for E2EE encryption, as well as about tag line information and profile pictures if available.
The collected dataset is highly sensitive, both locally and globally.
On a local level, it includes accounts from countries such as Myanmar, China, and North Korea --- nations where WhatsApp is officially banned and users risk severe penalties for its use.
On a global scale, the dataset represents an unprecedented level of exposure, containing more distinct users than any known leak and covering a significant portion of the global population.
If accessed by malicious actors, this data could be exploited for spam campaigns, phishing attacks, or robocalls, posing serious privacy and security risks.

By a census of the WhatsApp population, we show the extensive information that can be drawn from the dataset, including, among others, accounts per country and capita, growth and churn, the presence of companion devices, about text, and profile pictures, OS, device age, as well as account activeness and chattiness.
Beyond that, we identified around 2.3\,M distinct cryptographic (X25519) public keys which have been re-used in more than one device, including outstanding cases where certain public keys are reused multiple thousand times. 
Given the lack of plausible legitimate use cases, this behavior raises suspicion of abuse.
Finally, a comparison between the 2021 Facebook scraping incident and our dataset highlights the longevity of data leaks, as half of the phone numbers are still active.
As a countermeasure, we advise WhatsApp to (i) implement rate limiting, (ii) harmonize client implementations, (iii) encrypt profile information,  and (iv) %
better monitor for possibly fraudulent activity.
An even more desirable solution would be a widely adopted, federated instant messaging protocol standard that supports end-to-end encryption and mitigates risks associated with centralization --- such as data silos, vendor lock-in, and the ``all-eggs-in-one-basket'' problem.

\section*{Ethics Considerations}
\label{sec:ethical}

We recognize that phone number enumeration on a global scale is a privacy-sensitive issue. 
Due to the impact of such data leaks on users, instant messaging platforms, such as WhatsApp, are responsible for implementing safeguards against such enumeration attacks. 
Following the authors' approach in~\cite{hagen_2021_all}, we argue that empirical testing is the most reliable method to assess the feasibility and effectiveness of phone number enumeration in instant messaging applications. 
The authors highlight that this line of action is the only way to obtain a realistic estimate of the success of such attacks in real-world conditions.

\parvspace
\noindent
\textbf{Impact on Infrastructure and End Users.}
For all experiments, we gradually increased the load and only continued if no side effects (blocked accounts or significantly increased latency) were recognized.
All experiments were conducted from a single physical machine and IP address.
The enumeration process itself was spread over an extended period and required minimal bandwidth.
The most bandwidth- and load-intensive task involved downloading profile pictures. We began with a small number of threads and gradually scaled up to 1,000 concurrent threads to handle bulk requests for a limited target set.
To further verify the absence of effective rate-limits, we briefly tested bursts with 10,000 threads successfully, confirming that Meta's server infrastructure is not the bottleneck.
To this day, our IP address has not been blocked from accessing the relevant services. All requests terminated at WhatsApp's servers and did not reach user devices. As such, our enumeration study had no direct impact on end users.
To ensure the protection of any personally identifiable information, the gathered dataset was securely deleted prior to this paper's public release.

\parvspace
\noindent
\textbf{Responsible Disclosure.}
As part our first disclosure process, initiated on September 5, 2024 (ticket \#10211566797802504) and before we started investigating enumeration via the corresponding endpoints ourself, we already had reported to Meta that their XMPP API endpoints seem to have little to no rate-limiting, leading to numerous security and privacy issues.
To mitigate these issues, the paper that was attached to the report~\cite{gegenhuber_2024_carelesswhisper} explicitly mentioned rate-limits as a countermeasure.
After an initial response on September 24, 2024 saying that the paper will be forwarded to the corresponding engineering team, we received no further update till August 8, 2025, again informing us that the paper will be forwarded to the relevant engineering teams.
Throughout the year, we informed them about the corresponding submissions, paper acceptance and a correlating DEF CON talk~\cite{defcon33_silent_signals_2025}.

Due to this limited interest by Meta, we proceeded with our investigations on the feasibility of phone number enumeration independently.
We expected robust mitigation efforts to significantly obstruct our attempts, but that assumption proved incorrect.
During the entire phone number enumeration of WhatsApp, we were neither contacted by Meta on our reported issue, nor via the abuse contact that was available for our IP address. We made sure to be identifiable as research unit (by our university IP and reverse DNS entry) and reachable via the corresponding abuse email contact. 

On March 28, 2025, we initiated a second responsible disclosure process with Meta (ticket \#10212619137590341, title 
\emph{``Missing Rate Limiting in WhatsApp's API Causing Security and Privacy Issues''}).
In this second ticket, highlighting missing rate-limits in the XMPP endpoint for prekey retrieval~\cite{gegenhuber_2025_prekeypogo}, we also explicitly reported the potential for phone number enumeration, emphasizing its particular risk to users in countries where WhatsApp is officially banned, and urged Meta to implement a fix.
Two days later, the reported issue was flagged as a duplicate of the original one (from September 5th, 2024) and received no further attention. 

In a third attempt, on April 14, 2025 (ticket \#10212728845092960), we informed Meta that large-scale enumeration had indeed been possible.
We also mentioned that we were conducting a macroscopic analysis of the data as part of a scientific study and highlighted that we would neither publish the raw data nor retain it beyond completing our analysis.
Additionally, we raised the issue of recurring X25519 public keys, provided information on the most frequent keys, and asked for an explanation.
As a response, we received a general questionnaire on enumeration attacks (\emph{``What is the URL affected by the issue you found? Are you able to enumerate i) email used by users? ii) usernames used by users iii) phone numbers used by users?''}) that we answered.
Afterwards, the issue was marked as \emph{``not applicable''} and closed. 

Strongly believing that phone number enumeration at a large scale --- as highlighted in this paper --- is a serious issue,
we reopened the ticket on April 16, 2025, explicitly stating the enumeration speed achieved (i.e., 25\,M phone numbers per hour and session) to further underscore the criticality and potential impact of the vulnerability.
On April 17, 2025, we received the following response: \emph{``A member of Meta's security team has seen your report and performed an initial evaluation. We will get back to you once we have more information to share.''}

On May 29, 2025, we informed them about the second round advancement at NDSS and asked for an update, which again resulted in two generic responses:
\begin{enumerate}
    \item \emph{``I have forwarded your message to the appropriate points of contact, and we will follow up with you as soon as possible.''}
    \item \emph{``Thank you for your patience regarding your submissions. We apologise for the delays these reports encountered. I can confirm that the security teams have reviewed your research papers and forwarded some follow-ups we identified to the relevant engineering teams. We will update you as your report moves forward in our pipeline.''})
\end{enumerate}

On July 18, 2025, as part of the NDSS rebuttal, we confirmed that, large-scale enumeration is still possible (from the very same host and IP).

On August 9, 2025, we informed them of the NDSS acceptance and our plan to release a preprint by the end of the month, given that substantial time had already passed without any indication that Meta was actively addressing the issue.
While publishing before a fix carries the risk of enabling malicious actors to replicate the findings and extract data, we could not rule out that the vulnerability was already being exploited in the wild.
Since Meta at that point seemed unwilling to take action, the only remaining countermeasures were for users themselves, such as disabling their WhatsApp account or adjusting privacy settings.
Since uncoordinated publication of the preprint was considered a last resort, we explicitly stated that we were willing to provide additional time if steps were taken to address the issue.

On August 22, 2025, we reached out again to emphasize the urgency, attached a preprint of the current version of the paper, informing them that it would be published on September 4, 2025.

On August 30, 2025, they reached out to us, asking us to postpone the publication and invited us to discuss the paper in a conference call.
Throughout multiple meetings (September 2nd, September 5th, September 8th), they acknowledged our findings and we agreed on postponing the publication and actively assisting in their remediation efforts, which are still ongoing at the current date (September 10, 2025).

On September 10, 2025, we received an official statement from WhatsApp for the NDSS camera-ready version of this paper\footnote{\url{https://dx.doi.org/10.14722/ndss.2026.230805}}. A current version of this statement can be found at the bottom of the paper at hand.

\section*{Addendum (updated on 2025-11-11)}
Throughout September, October, and November 2025, we were in close communication with WhatsApp.
We provided them with additional technical details, including proof of concept code demonstrating user enumeration and the retrieval of associated account metadata, and discussed possible countermeasures.

In contrast to the lengthy process of establishing contact, WhatsApp took the matter seriously once we managed to make them aware of the identified issues.
According to WhatsApp, they did not assign CVEs for the discovered issues because they do not provide CVEs for server-side issues and state that the client-side bugs (e.g., key reuse, metadata leaks, OS disclosure) did not meet the threshold due to low impact.

We subsequently assisted them in retesting and confirming that the countermeasures had been rolled out and were effective by early October 2025.
Some of the countermeasures were implemented as a direct result of our collaboration (e.g., key reuse, profile picture timestamp), while others were already in development prior to our disclosure but had their implementation and rollout accelerated.

In more detail, the following countermeasures were implemented by WhatsApp:

\subsection{Cardinality Checks and Rate Limits}
In addition to broader rate and data limits adjustments, WhatsApp built cardinality counters to limit how many unique accounts you can query over an account's lifetime.
For context, naive strict limitations on the number or rate of queries you can perform would restrict normal operations as they do not distinguish between queries made to support ongoing interactions and querying a large number of distinct users.
In order to prevent scraping, it is necessary to estimate the cardinality of the data requested.
To do this in a privacy preserving manner WhatsApp has introduced cardinality estimation using probabilistic data structures~\cite{Janson2025HyperLogLog}.
These cardinality estimates are fundamental blocks for both upper limits on lifetime cardinality and within machine learning models to identify malicious activity at an early point in the lifecycle.
Using machine learning and cardinality counting, WhatsApp was able to effectively apply tailored data limits to likely scrapers while preserving regular application usage for users.

\subsection{Data Visibility Restrictions: Profile Picture and About Text}
WhatsApp began rolling out mitigations to limit the number of public profile pictures and about taglines a user can query over time. %
These restrictions apply to all cases – even when users decide to set their profile picture or about visibility to \textit{``Everyone''} in their privacy setting.
This effectively limits the ability for scraping publicly available pictures and about taglines at scale.
These restrictions don't apply to business accounts' profile picture to help businesses be recognized and build trust with their customers on WhatsApp.

\subsection{Profile Picture Timestamps}
Retrieving profile pictures no longer returns a timestamp of when it was changed.

\subsection{Key Reuse} %
WhatsApp identified a corner case on Android clients related to logouts and phone number changes, which led to the omission of fresh key generation during subsequent account setups. %
This issue has been addressed in WhatsApp version 2.25.26.11 and later.

\bigskip

On November 8, 2025, we received the following official statement from WhatsApp for the extended version of this paper:
\emph{``We welcome the work of the bug bounty community that helps us improve how WhatsApp
protects information. In this study, academic researchers generated a list of phone numbers,
checked if they are registered on WhatsApp and compiled basic public information that people have
made available to “everyone” in a novel manner that exceeded our intended limits. We have rolled
out new mitigations, including some of our industry’s leading anti-scraping systems we’d been
already working on prior to this study. We’re grateful to the researchers for their collaboration on
mitigation testing and hardening our defenses as a result. As a reminder, user messages remain
private and secure thanks to WhatsApp’s default end-to-end encryption. Finally, Meta’s Bug Bounty
team is working to improve our internal triage process to ensure that we more promptly respond to
academic submissions involving additional complexity.''}

\section*{Acknowledgments}

We would like to thank Markus Maier for his outstanding support in providing and maintaining the IT infrastructure, which was essential for the success of this work.

We also want to thank our anonymous reviewers for their valuable feedback and suggestions.

This material is based upon work partially supported by
(1) the University of Vienna, Faculty of Computer Science, Security \& Privacy Group,
(2) the University of Vienna, Faculty of Computer Science, Communication Technologies Group,
(3) the FFG Bridge project 46322124 SecKey, 
(4) the FFG KIRAS/K-PASS project 59103683 TelCrit,
(5) the Austrian Science Fund (FWF) (SFB SPyCoDe F85),
(6) SBA Research (SBA-K1 NGC) is a COMET Center within the COMET – Competence Centers for Excellent Technologies Programme and funded by BMIMI, BMWET, and the federal state of Vienna. The COMET Programme is managed by FFG.

\bibliographystyle{IEEEtran}
\bibliography{rfc, bibliography, messaging}

\appendix
\section{Data}

\begin{table*}%
\centering
\newcommand{\rotl}[1]{\multicolumn{1}{c}{\rlap{\rotatebox{60}{#1}}}}
{\rowcolors{3}{lightgray!30}{}
\begin{adjustbox}{height=0.48\textheight}
\begin{tabular}{llrr||rrrrrrrr}
\multicolumn{3}{c}{} & \rotl{} & \rotl{Accounts per 100 Capita} & \rotl{Android} & \rotl{iOS} & \rotl{Public Profile Picture} & \rotl{Public About Text} & \rotl{Business Account} & \rotl{Companion Device(s)} \\
\multicolumn{1}{c}{} & \multicolumn{1}{l}{Country} & \multicolumn{1}{r}{\# Accounts} & Global Share & \multicolumn{6}{c}{(\%)} \\
\midrule
1 & India & 749,075,246 & 21.67 & 52.3 & 95 & 5 & 62.2 & 29.5 & 9.8 & 6.2 \\
2 & Indonesia & 235,245,077 & 6.81 & 84.0 & 92 & 8 & 49.1 & 27.5 & 10.7 & 9.3 \\
3 & Brazil & 206,949,224 & 5.99 & 98.2 & 81 & 19 & 61.1 & 41.5 & 10.3 & 15.5 \\
4 & United States & 137,859,284 & 3.99 & 40.3 & 33 & 67 & 44.0 & 32.8 & 2.4 & 6.1 \\
5 & Russia & 132,855,022 & 3.84 & 91.2 & 76 & 24 & 61.7 & 33.5 & 3.6 & 9.4 \\
6 & Mexico & 128,324,166 & 3.71 & 99.3 & 82 & 18 & 46.1 & 23.3 & 4.1 & 11.7 \\
7 & Pakistan & 98,277,665 & 2.84 & 40.0 & 95 & 5 & 58.5 & 20.0 & 21.7 & 5.4 \\
8 & Germany & 74,565,425 & 2.16 & 88.3 & 58 & 42 & 51.0 & 35.4 & 2.2 & 13.4 \\
9 & Türkiye & 72,131,903 & 2.09 & 82.7 & 73 & 27 & 48.0 & 33.4 & 3.0 & 12.0 \\
10 & Egypt & 69,317,806 & 2.01 & 61.1 & 90 & 10 & 53.2 & 25.1 & 11.3 & 6.1 \\
11 & United Kingdom & 66,835,822 & 1.93 & 97.7 & 43 & 57 & 57.2 & 45.3 & 3.4 & 10.2 \\
12 & Iran & 64,945,192 & 1.88 & 72.1 & 76 & 24 & 53.6 & 23.1 & 9.1 & 1.5 \\
13 & Nigeria & 60,325,977 & 1.75 & 26.8 & 86 & 14 & 61.0 & 23.7 & 23.2 & 5.6 \\
14 & Bangladesh & 60,127,633 & 1.74 & 35.3 & 96 & 4 & 60.2 & 9.7 & 4.9 & 5.9 \\
15 & Italy & 55,606,677 & 1.61 & 93.4 & 70 & 30 & 68.5 & 46.9 & 3.1 & 16.5 \\
16 & France & 53,967,206 & 1.56 & 81.3 & 58 & 42 & 50.6 & 36.7 & 1.6 & 10.2 \\
17 & Colombia & 51,030,811 & 1.48 & 98.1 & 87 & 13 & 48.4 & 24.3 & 8.6 & 14.3 \\
18 & Spain & 46,495,245 & 1.35 & 97.1 & 72 & 28 & 58.2 & 42.2 & 3.8 & 18.6 \\
19 & Argentina & 43,854,434 & 1.27 & 96.5 & 88 & 12 & 54.8 & 32.8 & 5.4 & 16.2 \\
20 & South Africa & 43,171,922 & 1.25 & 68.7 & 90 & 10 & 59.0 & 25.6 & 10.3 & 5.2 \\
21 & Saudi Arabia & 38,910,335 & 1.13 & 118.4 & 57 & 43 & 61.3 & 38.9 & 16.0 & 8.4 \\
22 & Malaysia & 37,353,780 & 1.08 & 107.0 & 77 & 23 & 50.8 & 29.6 & 7.9 & 15.4 \\
23 & Morocco & 32,954,817 & 0.95 & 87.8 & 89 & 11 & 54.4 & 17.6 & 15.6 & 5.6 \\
24 & Peru & 31,567,678 & 0.91 & 93.8 & 93 & 7 & 49.0 & 25.1 & 7.2 & 19.7 \\
25 & Iraq & 29,676,773 & 0.86 & 66.6 & 79 & 21 & 67.6 & 29.9 & 7.9 & 2.5 \\
26 & Poland & 23,838,052 & 0.69 & 61.4 & 75 & 25 & 43.5 & 25.5 & 0.8 & 7.7 \\
27 & Canada & 21,978,575 & 0.64 & 56.3 & 37 & 63 & 44.1 & 40.4 & 4.5 & 11.1 \\
28 & Philippines & 21,393,429 & 0.62 & 18.7 & 82 & 18 & 36.2 & 6.4 & 1.9 & 3.3 \\
29 & Kenya & 21,094,162 & 0.61 & 38.5 & 97 & 3 & 60.6 & 23.4 & 15.4 & 4.6 \\
30 & Chile & 21,020,792 & 0.61 & 107.2 & 80 & 20 & 52.8 & 30.8 & 5.1 & 18.5 \\
31 & Kazakhstan & 19,702,340 & 0.57 & 97.6 & 71 & 29 & 59.0 & 37.2 & 10.0 & 10.7 \\
32 & Netherlands & 18,668,743 & 0.54 & 103.6 & 51 & 49 & 59.3 & 45.8 & 4.5 & 22.4 \\
33 & Venezuela & 18,645,768 & 0.54 & 66.0 & 94 & 6 & 52.0 & 24.0 & 9.3 & 5.7 \\
34 & UAE & 18,348,803 & 0.53 & 175.7 & 69 & 31 & 61.9 & 31.1 & 26.3 & 12.2 \\
35 & Algeria & 18,316,732 & 0.53 & 40.0 & 92 & 8 & 42.7 & 6.1 & 2.3 & 1.8 \\
36 & Ghana & 16,671,248 & 0.48 & 49.8 & 78 & 22 & 69.8 & 28.3 & 23.7 & 4.0 \\
37 & Australia & 16,349,836 & 0.47 & 62.1 & 33 & 67 & 44.5 & 42.3 & 1.8 & 9.3 \\
38 & Ecuador & 16,024,559 & 0.46 & 89.5 & 88 & 12 & 48.2 & 25.0 & 9.5 & 18.1 \\
39 & Romania & 15,081,156 & 0.44 & 78.7 & 78 & 22 & 62.0 & 29.4 & 2.1 & 14.2 \\
40 & Sri Lanka & 14,899,815 & 0.43 & 65.1 & 88 & 12 & 52.3 & 23.5 & 13.9 & 9.0 \\
41 & Ukraine & 14,261,123 & 0.41 & 37.5 & 71 & 29 & 42.3 & 19.9 & 1.0 & 6.4 \\
42 & Guatemala & 13,402,143 & 0.39 & 74.5 & 91 & 9 & 47.6 & 20.1 & 6.0 & 8.2 \\
43 & Nepal & 13,110,193 & 0.38 & 44.1 & 90 & 10 & 54.3 & 10.6 & 3.9 & 4.7 \\
44 & Syria & 12,515,809 & 0.36 & 54.4 & 96 & 4 & 58.1 & 23.9 & 14.2 & 2.0 \\
45 & Yemen & 12,444,545 & 0.36 & 32.1 & 97 & 3 & 63.5 & 24.4 & 18.4 & 1.4 \\
46 & Tanzania & 12,337,074 & 0.36 & 18.8 & 93 & 7 & 70.8 & 20.6 & 17.1 & 2.3 \\
47 & DR Congo & 11,846,557 & 0.34 & 11.4 & 91 & 9 & 77.5 & 16.9 & 29.5 & 1.6 \\
48 & Israel & 11,273,740 & 0.33 & 122.7 & 64 & 36 & 63.8 & 45.1 & 6.3 & 22.3 \\
49 & Senegal & 11,185,047 & 0.32 & 62.6 & 83 & 17 & 75.4 & 22.3 & 20.3 & 2.7 \\
50 & Uganda & 10,723,769 & 0.31 & 22.4 & 95 & 5 & 73.7 & 23.5 & 23.5 & 2.4 \\
51--245 & & 430,063,259 & 12.44 & 14.3 & 77 & 23 & 58.2 & 25.6 & 10.3 & 7.6 \\
\midrule
\multicolumn{2}{l}{\textbf{GLOBAL (245 countries)}} & 3,456,622,389 & 100.00 & 42.9 & 81 & 19 & 56.7 & 29.3 & 9.0 & 8.8 \\
\bottomrule
\end{tabular}
\end{adjustbox}
}
\vspace{1ex}
\caption{Extended version of Figure~\ref{tab:result-overview-top-10}.}
\label{tab:result-overview-top-50}
\end{table*}

\begin{figure}[t]
    \centering
    \includegraphics[width=0.8\columnwidth]{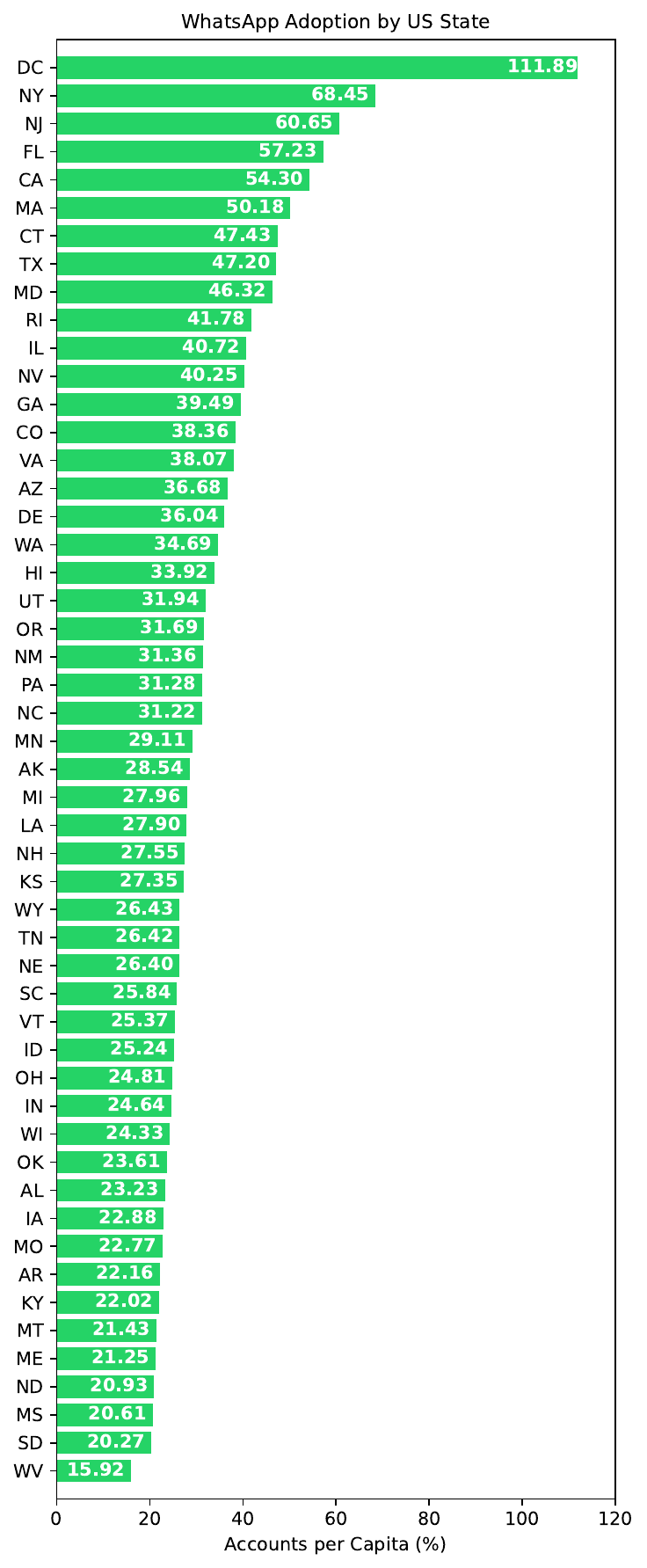}
    \vspace{-0.3cm}
    \caption{In some countries, such as the US, Brazil, and Mexico, phone numbers are assigned to specific geographical regions based on area codes.
    We used population statistics to calculate a fine-grained WhatsApp adoption for the US (accounts per capita).
    }
    \label{fig:market-penetration-us-states-chart}
    \vspace{-0.2cm}
\end{figure}

\begin{figure}[t]
  \centering
  \definecolor{whatsapp-green-light}{HTML}{25D366}
  \definecolor{whatsapp-green-teal}{HTML}{075E54}
  \includesvg[width=\linewidth]{figures/map_market_penetration_us_states.svg}
  {\footnotesize WhatsApp Accounts per Capita}\\
  {\footnotesize 0\,\%}~\tikz{\fill[left color=white,right color=whatsapp-green-teal] (0,0) rectangle (0.6\linewidth,0.5em);}~{\footnotesize 100\,\%}
  \caption{In some countries, such as the US, Brazil, and Mexico, phone numbers are assigned to specific geographical regions based on area codes.
    We used population statistics to calculate a fine-grained WhatsApp adoption for the US (accounts per capita).}
  \label{fig:market-penetration-us-states-map}
\end{figure}

\begin{figure}[b]
    \centering
    \includegraphics[width=\linewidth,trim={0.2cm 0.25cm 0.2cm 0.2cm},clip]{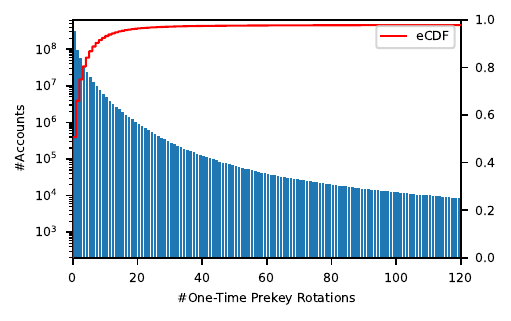}
    \vspace{-0.3cm}
    \caption{Since new conversations exhaust one-time prekeys and eventually lead to the upload of a fresh prekey batch (i.e., a prekey bundle rotation), older or more active iOS accounts tend to have higher prekey IDs. Since every batch contains 812 one-time prekeys, most devices only had several rotations --- this can be observed by the eCDF line.}
    \label{fig:device-chattiness}
    \vspace{-0.2cm}
\end{figure}

\begin{figure*}[t]
    \centering
    \includegraphics[width=\textwidth]{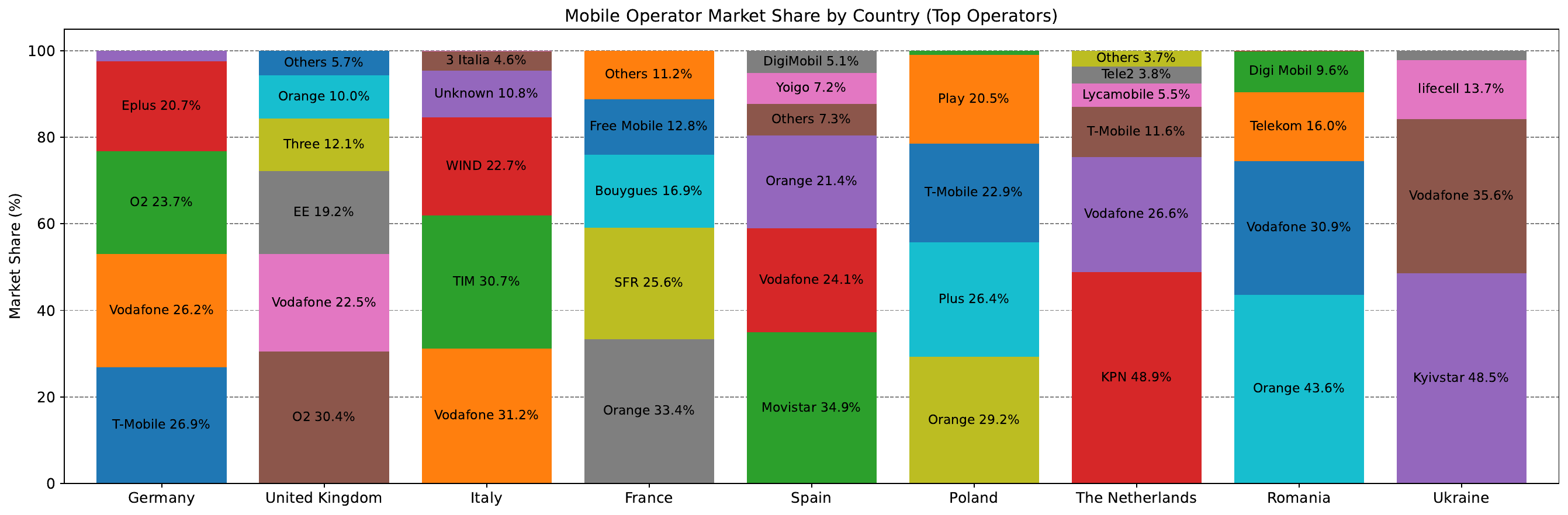}
    \vspace{-0.7cm}
    \caption{Operator market share as classified by \textit{libphonenumber} for active WhatsApp numbers discovered in different target countries.
    While these values provide an estimate of an operator's market share, they are subject to several limitations, such as not accounting for number porting.
    }
    \vspace{-0.3cm}
    \label{fig:market-share-operators}    
\end{figure*}

\begin{figure}[t]
  \centering
  \includegraphics[width=0.9\linewidth]{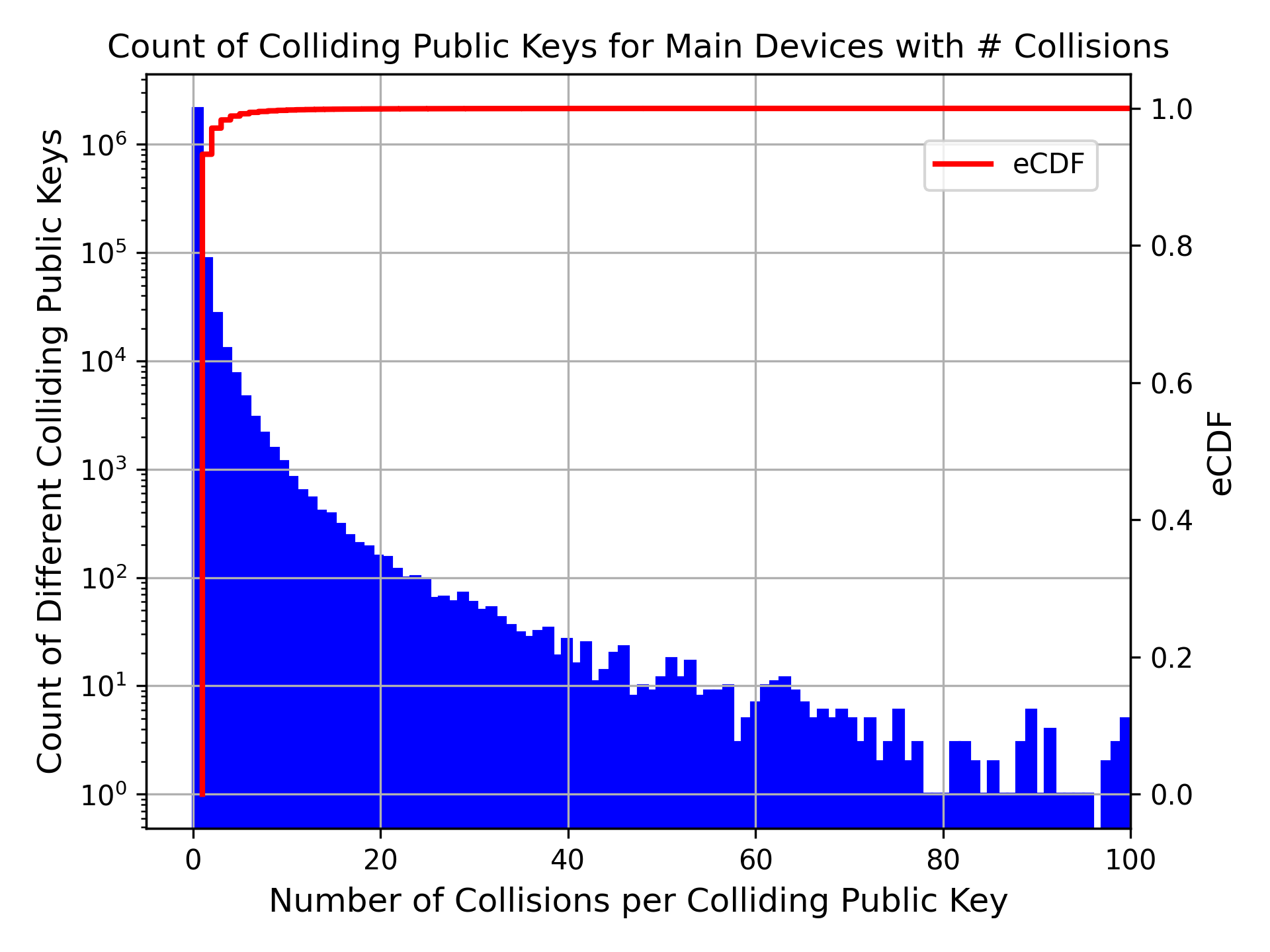}
  \vspace{-0.3cm}
  \caption{Histogram showing counts for different colliding public keys (y axis log scale) of \emph{main} devices (device index $= 0$), with their respective number of collisions per public key (x axis). The figure is truncated by 100 collisions and thus outlier with high numbers of collisions are not displayed. See Table~\ref{tab:top_collisions} for public keys with the largest number of occurrences.}
  \label{fig:hist_coll_count_100}
  \vspace{-0.1cm}
\end{figure}

\begin{figure}[t]%
  \centering
  \includegraphics[width=0.9\linewidth]{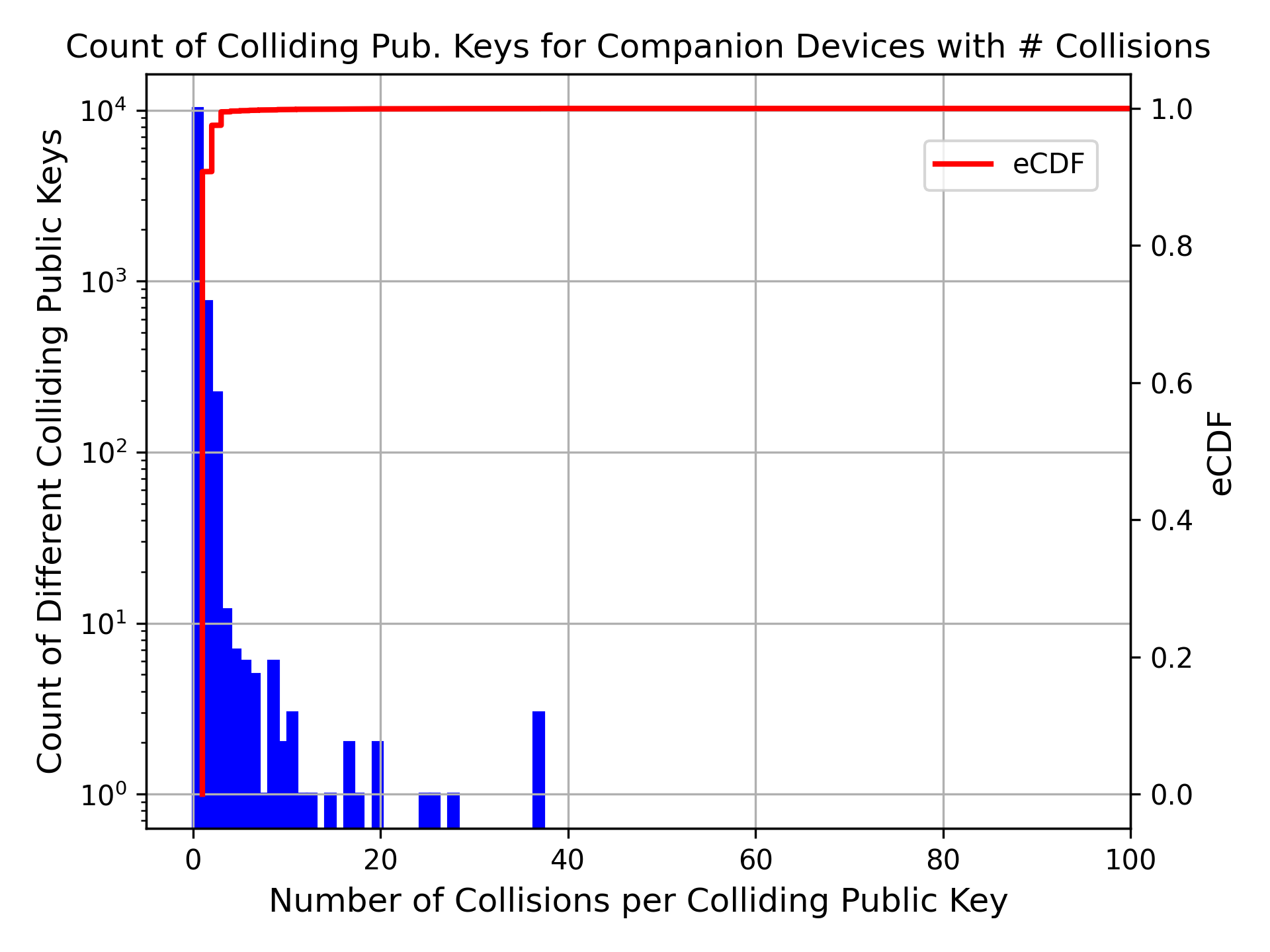}
  \vspace{-0.3cm}
  \caption{Histogram showing counts for different colliding public keys (y axis log scale) of \emph{companion} devices (device index $> 0$), with their respective number of collisions per public key (x axis). The figure is truncated by 100 collisions and thus outlier with high numbers of collisions are not displayed. See Table~\ref{tab:top_collisions} for public keys with the largest number of occurrences.}
  \label{fig:hist_coll_count_companion_100}
  \vspace{-0.1cm}
\end{figure}

\begin{figure}[t]%
  \centering
  \includegraphics[width=0.99\linewidth]{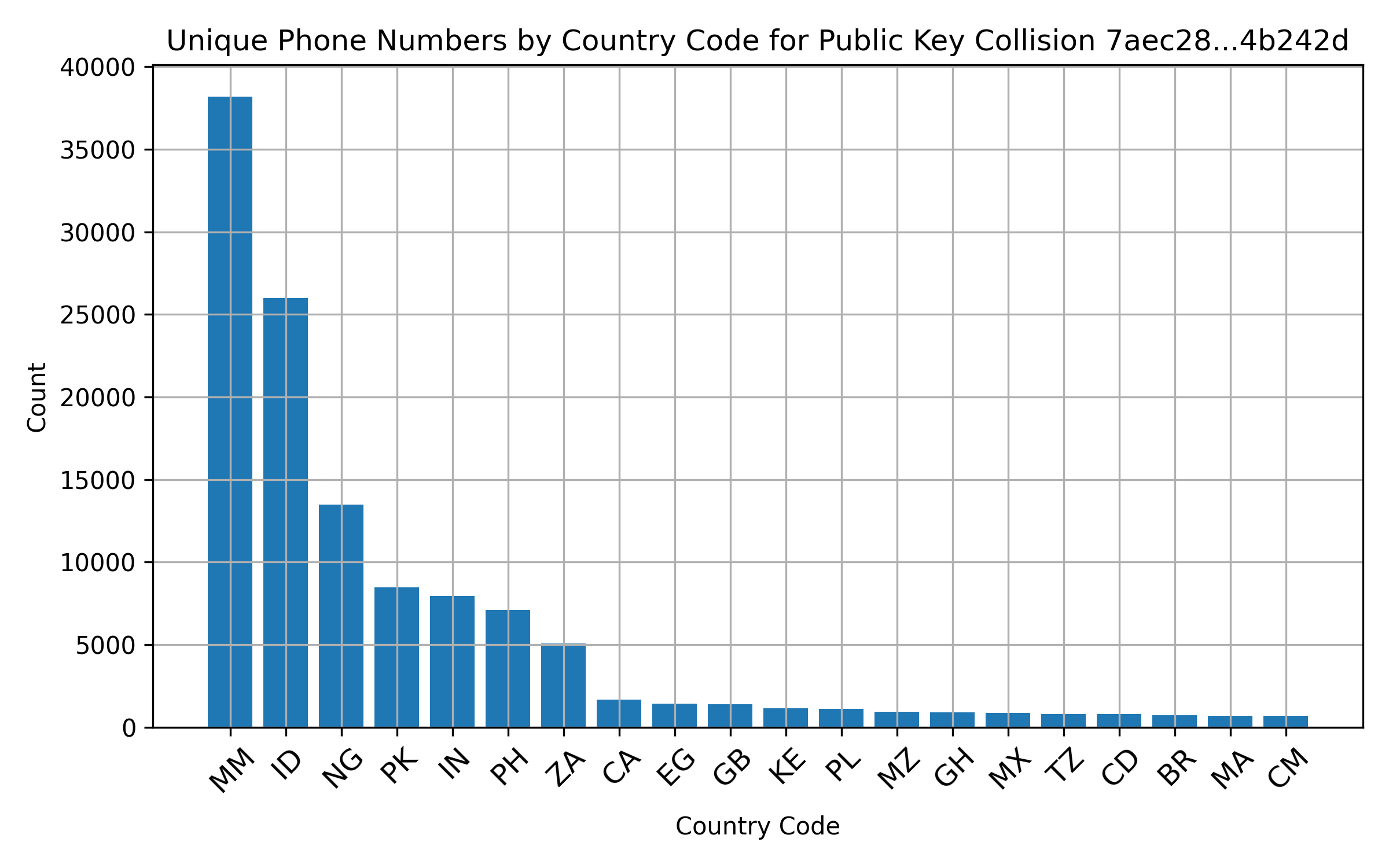}
  \vspace{-0.8cm}
  \caption{Phone numbers per country using the public key \texttt{7aec...} with the most number of occurrences in our dataset. Most of the phone numbers of affected devices can be attributed to Myanmar (MM) and Indonesia (ND) followed by Nigeria (NG).}
  \label{fig:country_code_hist_7ac28}
  \vspace{-0.2cm}
\end{figure}

\pagebreak

\end{document}